\documentclass[reprint, longbibliography, superscriptaddress, preprintnumbers, nofootinbib, nobibnotes, amsmath, amssymb, aps, pra, floatfix]{revtex4-2}

\usepackage[normalem]{ulem}
\usepackage{gensymb, upgreek, mathrsfs, comment, graphicx, dcolumn, bm, float, url, xcolor, textcomp, amsmath, varwidth, placeins, gensymb, siunitx, multirow}
\usepackage{amsfonts, amssymb, verbatim, dsfont, makecell, array, xspace, xifthen, esvect}

\usepackage[colorlinks=true, urlcolor=blue, linkcolor=blue, citecolor=blue]{hyperref}
\hypersetup{colorlinks, linkcolor={red!50!black}, citecolor={blue!50!black}, urlcolor={blue!80!black}}

\usepackage[center]{titlesec}
\titleformat{\section}{\centering\normalfont\fontsize{10}{10}\bfseries}{\thesection}{1em}{}
\titlespacing{\section}{0pt}{12pt plus 4pt minus 2pt}{6pt plus 2pt minus 2pt}
\titleformat{\subsection}{\centering\normalfont\fontsize{10}{10}\bfseries}{\thesubsection}{1em}{}
\titlespacing{\subsection}{0pt}{12pt plus 4pt minus 2pt}{6pt plus 2pt minus 2pt}

\usepackage{tabularx}
\newcolumntype{C}{>{\centering\arraybackslash}X}

\newcommand{\pP}[1]{\left(#1\right)}
\newcommand{\ro}[1]{\bm{#1}}

\newcommand{\pmat}[2][]{\ifthenelse{\equal{#1}{}}{\begin{pmatrix}#2\end{pmatrix}}{\begin{pmatrix}[#1]#2\end{pmatrix}}}

\newcommand{\bra}[1]{\ensuremath{\langle#1|}}
\newcommand{\ket}[1]{\ensuremath{|#1\rangle}}

\newcommand{\bracket}[3]{\ensuremath{\langle#1|#2|#3\rangle}}

\begin{document}

\title{Earth's field diamond vector magnetometry with isotropic magnetic flux concentrators}
\date{\today}

\author{Maziar~Saleh~Ziabari}

\affiliation{Center for High Technology Materials, 
University of New Mexico, Albuquerque, NM, USA}
\affiliation{Department of Physics and Astronomy,
University of New Mexico, Albuquerque, NM, USA}

\author{Nazanin~Mosavian}
\affiliation{Center for High Technology Materials, 
University of New Mexico, Albuquerque, NM, USA}
\affiliation{Department of Physics and Astronomy,
University of New Mexico, Albuquerque, NM, USA}

\author{Ilja~Fescenko}
\affiliation{Center for High Technology Materials, 
University of New Mexico, Albuquerque, NM, USA}
\affiliation{Laser Centre, Faculty of Science and Technology, University of Latvia, Riga, Latvia}

\author{Yaser~Silani}
\affiliation{Center for High Technology Materials, 
University of New Mexico, Albuquerque, NM, USA}

\author{Bryan~A.~Richards}
\affiliation{Center for High Technology Materials, 
University of New Mexico, Albuquerque, NM, USA}
\affiliation{Department of Physics and Astronomy,
University of New Mexico, Albuquerque, NM, USA}

\author{Andris~Berzins}
\affiliation{Center for High Technology Materials, 
University of New Mexico, Albuquerque, NM, USA}

\author{Maxwell~D.~Aiello}
\affiliation{Center for High Technology Materials, 
University of New Mexico, Albuquerque, NM, USA}
\affiliation{Department of Physics and Astronomy,
University of New Mexico, Albuquerque, NM, USA}

\author{Keith~A.~Lidke}
\affiliation{Department of Physics and Astronomy,
University of New Mexico, Albuquerque, NM, USA}

\author{Andrey~Jarmola}
\affiliation{ODMR Technologies Inc., El Cerrito, CA, United States}
\affiliation{University of California-Berkeley, Berkeley, CA, United States}

\author{Janis~Smits}
\affiliation{Center for High Technology Materials, 
University of New Mexico, Albuquerque, NM, USA}

\author{Victor~M.~Acosta}
\email{victormarcelacosta@gmail.com}
\affiliation{Center for High Technology Materials, 
University of New Mexico, Albuquerque, NM, USA}
\affiliation{Department of Physics and Astronomy,
University of New Mexico, Albuquerque, NM, USA}

\date{\today}

\begin{abstract}
    Vector magnetometers based on the optically detected magnetic resonance (ODMR) of nitrogen-vacancy centers in diamond are being developed for applications such as navigation and geomagnetism. However, at low magnetic fields, such as that on Earth (${\sim}50~{\rm \upmu T}$), diamond magnetometers suffer from spectral congestion whereby ODMR peaks are not easily resolved. Here, we experimentally investigate a potential solution of using an isotropic, three-dimensional magnetic flux concentrator to amplify Earth's field without altering its direction. The concentrator consists of six ferrite cones, in a face-centered cubic arrangement, centered about a diamond. We vary the direction of a $50~{\rm \upmu T}$ applied field and record and fit the resulting ODMR spectra. By comparing the fitted fields to those of a reference fluxgate magnetometer, we characterize the angular response of the diamond magnetometer and quantify absolute errors in the field magnitude and angle. We find that the enhancement factor is nearly isotropic, with a mean of $19.05$ and a standard deviation of $0.16$, when weighted by solid angle coverage. Gradient broadening of the ODMR lines is sufficiently small that the spectra are well resolved for nearly all field directions, alleviating spectral congestion. For ${\sim}98\%$ of the total $4\pi$ solid angle, Cram{\`e}r-Rao lower bounds for magnetic field estimation uncertainty are within a factor of $2$ of those of the fully-resolved case, indicating minimal deadzones. We track the stability of the magnetometer over six hours and observe variations $\lesssim40~{\rm nT/hour}$, limited by temperature drift. Our study presents a new route for diamond vector magnetometry at Earth's field, with potential applications in geomagnetic surveys, anomaly detection, and navigation.
\end{abstract}

\maketitle

\section{\label{sec:intro}Introduction}
Vector magnetometry in Earth's ambient magnetic field (${\sim}\,50~{\rm \upmu T}$) is important for applications such as navigation~\cite{AFZ2011,SHO2014, CAN2016, HUA2019}, mineral exploration~\cite{SCH2004,COL2005}, landmine detection~\cite{ZHA2003,YOO2021}, and geomagnetic surveys~\cite{VIN1963, NEU2001}.  A number of instruments are available for this purpose, including fluxgate magnetometers~\cite{PRI1979, BAS2010}, alkali-metal vapor magnetometers~\cite{LEE2021, BLO1960}, superconducting quantum interference devices~\cite{BIC1999, SCH2004}, Hall-effect probes~\cite{ROU2001, CHA2014}, and magnetoresistive sensors~\cite{GIA2012, BRO2014}. Recently, vector magnetometers based on the optically detected magnetic resonance (ODMR) of diamond nitrogen-vacancy (NV) centers have emerged as a promising alternative. Diamond magnetometers have reached sub-picotesla sensitivity~\cite{FES2020,GAO2023,BAR2024}, and they offer the advantage of simultaneously measuring all magnetic field vector components at a single position~\cite{MAE2010,SCH2018,ZHA2018}, owing to the four fixed NV crystallographic axes. They have been integrated into compact sensors~\cite{FRO2018,STU2019,WAN2023,MAO2023,SON2023,RAN2023,GRA2025} and tested in field applications~\cite{FRO2018,GRA2025}. 

However, realizing high accuracy in Earth's magnetic field remains a challenge for diamond vector magnetometers, due to spectral congestion. For a field magnitude of ${\sim}\,50~{\rm \upmu T}$, the 24 NV spin resonance frequencies are clustered over a range of ${\lesssim}\,7~{\rm MHz}$. For typical NV ODMR linewidths, $\Gamma\,{\approx}\,1~{\rm MHz}$ full-width-at-half-maximum (FWHM), the resonance frequencies cannot be uniquely resolved, preventing accurate vector measurements. Moreover, spectral shifts due to strain can lead to significant absolute errors at low fields~\cite{DOL2011,FAN2013,GLE2017}. 

A common solution to the spectral congestion problem is to apply a large bias field and measure Earth's field as a perturbation~\cite{SCH2018,FRO2018,GRA2025}. However, this method has the drawback of requiring a very stable bias field. For example, realizing sub-picotesla drift with a $1~{\rm mT}$ bias field requires ${\lesssim}\,1$ part-per-billion field stability. An alternative solution is to make use of the polarization anisotropy of the NV microwave and optical transitions to selectively address subsets of ODMR peaks~\cite{ALE2007,MRO2015,ZHE2019,MUN2020,CHI2025}, but this does not completely remove ambiguity due to imperfect polarization selection rules~\cite{ALE2007} and the significant contribution of strain shifts.

Magnetic flux concentrators have recently been integrated with diamond magnetometers to improve sensitivity by amplifying magnetic fields in either one~\cite{FES2020,SIL2023,MAO2023,GAO2023,SHA2023} or two~\cite{WAN2023} dimensions. We hypothesized that an \textit{isotropic} flux concentrator--one that amplifies all magnetic vector components equally--could both improve sensitivity and resolve the spectral congestion problem by amplifying Earth's field. Prior work combined triaxial flux concentrators with magnetoresistive~\cite{CHE2012,ZHA2013} and coil-based~\cite{LU2015,CHE2024} magnetometers, but the response was highly anisotropic~\cite{LU2015,CHE2024}.

\begin{figure*}[bt]
    \includegraphics[width=0.96\textwidth]{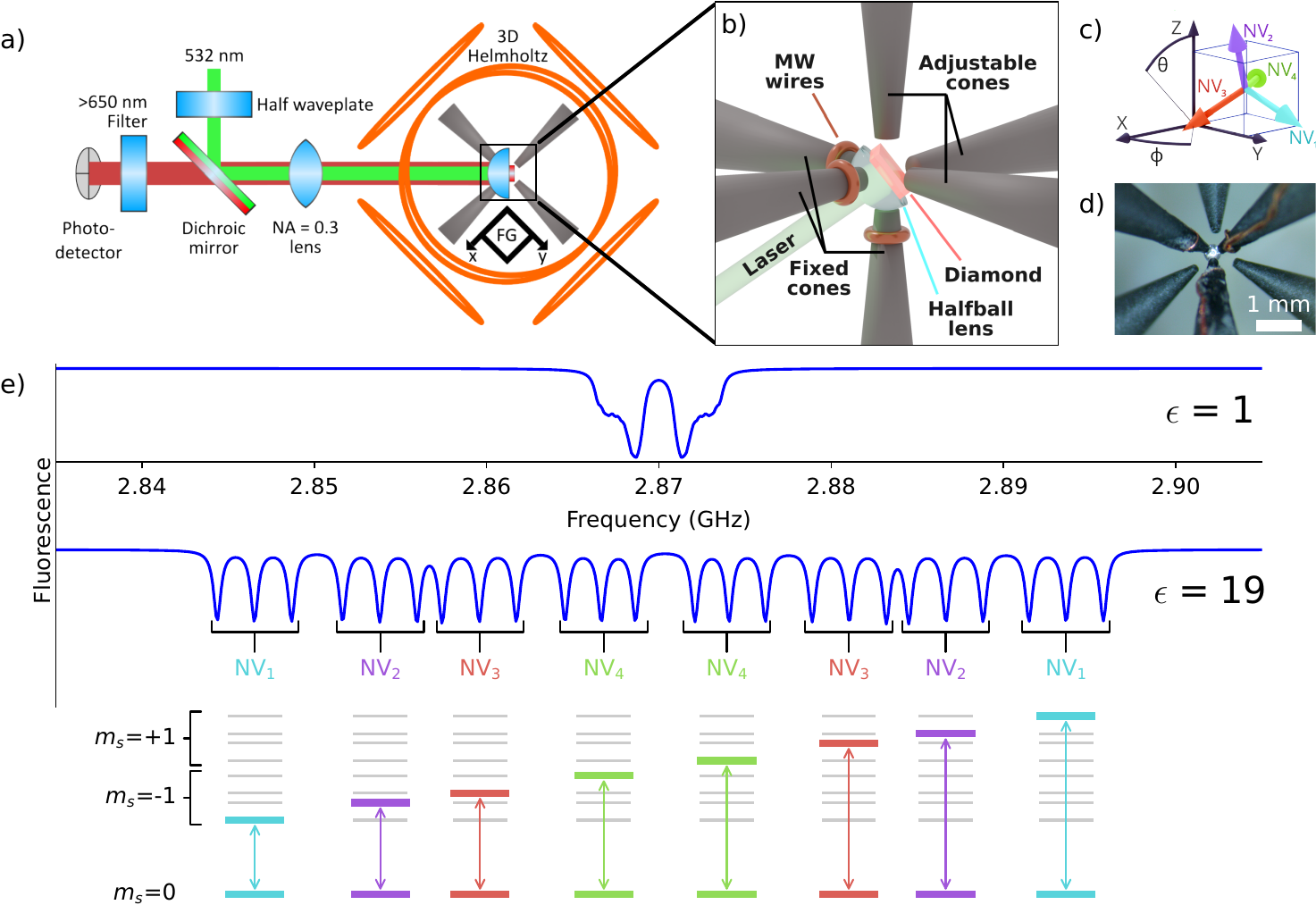}
    \caption{\textbf{Apparatus and operating principle.} (a) Diagram of the experimental setup. An epifluorescence microscope is used to perform NV ODMR spectroscopy. An adjustable polarizer on the NV fluorescence path is not shown. A three-axis Helmholtz coil assembly mimics Earth's magnetic field with arbitrary direction. A triaxial fluxgate magnetometer (FG) serves as a ``ground-truth'' reference. (b) Zoom-in of the isotropic flux concentrator, diamond membrane, and half-ball lens. Six MN60 ferrite cones (each $15~{\rm mm}$ long, with $6{\mbox{-}}{\rm mm}$-diameter base and $0.24{\mbox{-}}{\rm mm}$-diameter flat tip, see~\ref{app:assembly}) in a face-centered cubic configuration are positioned with tips ${\sim}0.3~{\rm mm}$ from the illuminated diamond volume. Three orthogonal cones are axially adjustable, while the others are affixed to the half-ball lens. Separate wires are wound around the three fixed cone tips to provide polarized microwaves (MW) (\ref{app:save_setup}). 
    (c) Position of the four NV axes with respect to the reference fluxgate Cartesian axes (``lab frame''). (d)  Photo of the assembly from the point of view of the incoming laser.  (e) Simulated ODMR spectrum at $|\vec{B}_{\rm applied}|=50~{\rm \upmu T}$ of NV centers without ($\epsilon=1$, top) and with ($\epsilon=19$, bottom) isotropic flux concentrators. The linewidth is chosen to be $\Gamma=0.5~{\rm MHz}$ FWHM. A diagram of the NV transitions associated with each of the NV axes and each of the $\Delta m_s=\pm1$ transitions is illustrated at the bottom ($^{14}$N hyperfine splitting is omitted for visual clarity). 
    }
    \label{fig:background}
\end{figure*}

Here, we experimentally demonstrate that a device consisting of three orthogonal pairs of ferrite cones acts as a nearly isotropic magnetic flux concentrator for diamond vector magnetometry. We vary the direction of an applied $50~{\rm \upmu T}$ field and find that ODMR spectra are well resolved for nearly all field directions. We characterize the angular response of the diamond magnetometer and quantify absolute errors in the field magnitude and angle. We apply a Cram{\`e}r-Rao bound analysis to evaluate angular ``deadzones'' and find that high-precision magnetic vector estimation is possible for approximately $98\%$ of the full $4\pi$ solid angle. Finally, we record the stability of the magnetometer and identify a potential source of systematic error due to temperature drift. Our study reveals the promise of isotropic flux concentrators for Earth's field diamond vector magnetometry.

\section{\label{sec:experiment}Experiment}
The experimental apparatus is depicted in Fig.~\ref{fig:background}(a), with additional details in \ref{app:setup}. An epifluorescence configuration is used to perform NV ODMR spectroscopy. Light from a $532~{\rm nm}$ laser (${\sim}10~{\rm mW}$) is focused by a plano-convex lens (numerical aperture, NA$=0.3$) and sapphire half-ball lens onto an NV-doped diamond. NV fluorescence is collected by the same lenses, spectrally filtered, and focused onto a photodetector. Three orthogonal pairs of Helmholtz coils are used to zero the ambient lab field and apply a magnetic field in any direction. A triaxial fluxgate magnetometer, near the center of the coils, serves as a reference to record the applied magnetic field vector and defines the ``lab frame''.

Figure~\ref{fig:background}(b) illustrates the miniature components of the apparatus. A $^{12}$C isotopically purified diamond, doped with ${\sim}4~{\rm ppm}$ of NV centers, is cut into dimensions ${\sim}250\times250\times50~{\rm \upmu m^3}$, with the large faces polished normal to a [110] crystallographic direction~\cite{JAR2021,SMI2025}. The membrane is affixed to a $500{\mbox{-}}{\rm \upmu m}$-diameter sapphire half-ball lens and placed at the center of an isotropic magnetic flux concentrator. The flux concentrator consists of three orthogonally aligned MN60 ferrite cone pairs, with the NV illumination volume at the center. The cones are mounted within a machined polyetheretherketone (PEEK) housing. One cone from each orthogonal pair is affixed to the half-ball lens, while the other three cones can be translated along their cone axes, using fine-thread screws, with micrometer accuracy. Copper wires wound around each of the three fixed cones are used to apply microwave fields with adjustable polarization. Figure~\ref{fig:background}(c) illustrates the directions of the four NV axes with respect to the lab frame, as defined by the fluxgate magnetometer's axes. Figure~\ref{fig:background}(d) shows a photograph of the fabricated device. 
\ref{app:assembly} discusses the device design and fabrication.

Figure~\ref{fig:background}(e) shows simulated ODMR spectra, at $|\vec{B}_{\rm applied}|\,{=}\,50~{\rm \upmu T}$, with and without an isotropic flux concentrator of enhancement factor $\epsilon=19$. NV centers consist of an $S=1$ electronic spin, which is weakly coupled to an intrinsic $I=1$ $^{14}$N nuclear spin. Each NV center has six allowed spin transitions ($\Delta m_s=\pm1$, $\Delta m_i=0$). For an NV center aligned along the $\hat{z}$ direction, the ODMR frequencies are calculated by applying the spin transition selection rules after diagonalizing the Hamiltonian:
\begin{align}
  \mathcal{H} \,\approx\,
  & D S_z^2 + E({S_x^2-S_y^2}) + \gamma_{\rm nv\,}\vec{B}_{\rm nv}\cdot\vec{S} \notag\\
  & + A_{||} S_zI_z + PI_z^2 + \gamma_{\rm nuc\,}B_{{\rm nv},z} I_z,
    \label{eqn:hamiltonian}
\end{align}
where $D\,{=}\,2.87~{\rm GHz}$ and $E\,{\approx}\,1~{\rm MHz}$ are the axial and transverse zero-field splitting parameters respectively, $\gamma_{\rm nv}\,{=}\,28.03~{\rm GHz/T}$ is the NV gyromagnetic ratio, $\vec{B}_{\rm nv}$ is the magnetic field vector within the NV illumination region, $A_{||}\,{=}\,-2.17~{\rm MHz}$ is the axial hyperfine constant, $P\,{=}\,-4.95~{\rm MHz}$ is the quadrupole splitting parameter, and $\gamma_{\rm nuc}\,{=}\,-3.07~{\rm MHz/T}$ is the $^{14}$N nuclear gyromagnetic ratio. Note that Eq.~\eqref{eqn:hamiltonian} omits transverse hyperfine and nuclear Zeeman terms, which have negligible contribution for the low fields studied here.

There are four possible NV axes, which we denote by unit vector $\hat{e}_{\kappa}$. Each NV axis is assumed to be equally populated in the diamond and has its own set of six spin transitions constrained by Eq.~\eqref{eqn:hamiltonian}. The resulting ODMR spectrum thus has a total of 24 peaks. For small magnetic fields $\gamma_{\rm nv} |\vec{B}_{\rm nv}|\ll D$, the 24 ODMR frequencies are approximately given by:
\begin{equation}
    \label{eqn:appxenergy}
    f^{\pm}_{i,\kappa} \approx D\pm(\gamma_{\rm nv} \vec{B}_{\rm nv}\cdot\hat{e}_{\kappa}+A_{||}m_i).
\end{equation}
 
 As seen from Eq.~\eqref{eqn:appxenergy}, the 24 ODMR frequencies are spread over a range of at most $2(\gamma_{\rm nv} |\vec{B}_{\rm nv}|+A_{||})\approx7~{\rm MHz}$ at $|\vec{B}_{\rm nv}|\approx50~{\rm \upmu T}$. An example simulated spectrum is plotted at the top of Fig.~\ref{fig:background}(e) for the case of no flux concentrator with $|\vec{B}_{\rm nv}|=|\vec{B}_{\rm applied}|=50~{\rm \upmu T}$ and Lorentzian ODMR FWHM linewidths set to a relatively-low value, $\Gamma=0.5~{\rm MHz}$. Here, there is significant spectral congestion. All 24 ODMR peaks overlap, and it is difficult to uniquely determine the magnetic field vector. With an isotropic magnetic flux concentrator, the internal field within the diamond is amplified, $\vec{B}_{\rm nv}=\epsilon\vec{B}_{\rm applied}$. In Fig.~\ref{fig:background}(e), an example simulated ODMR spectrum is plotted for the same $\vec{B}_{\rm applied}$ and $\Gamma=0.5~{\rm MHz}$, but now with $\epsilon=19$. All 24 ODMR peaks are well resolved for this field orientation, allowing for accurate estimation of $\vec{B}_{\rm nv}$ and, for a known enhancement factor, $\vec{B}_{\rm applied}$ (\ref{app:broadening}).

\section{Isotropic flux concentrator design} 
\label{sec:isotropic}
\begin{figure}[hbt]
     \includegraphics[width=0.95\columnwidth]{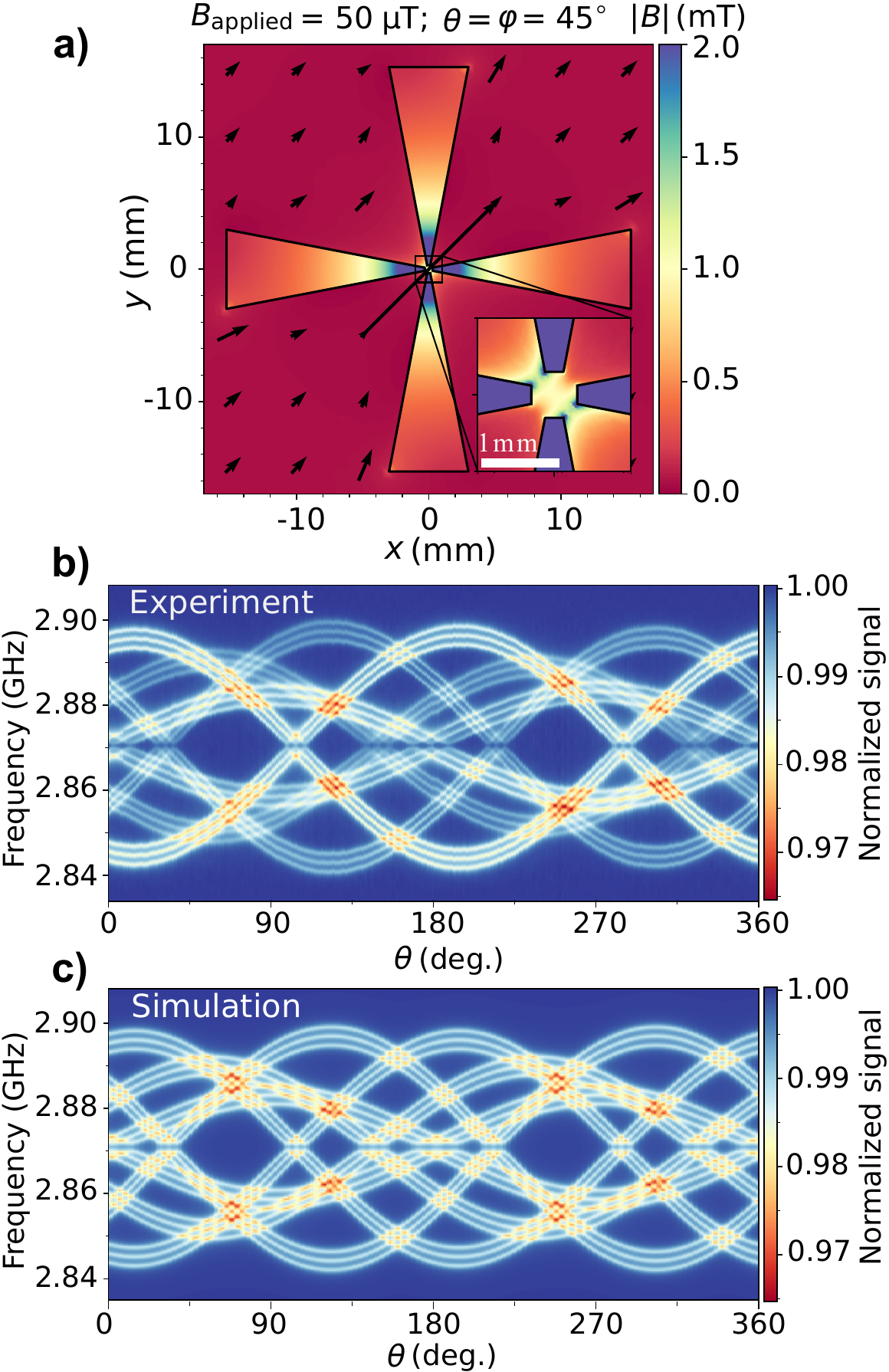}
     \caption{\textbf{Flux concentrator design and validation.} (a) $xy$-plane magnetic map of the device, simulated by three-dimensional finite-element magnetostatic modeling, for $|\vec{B}_{\rm applied}|=50~{\rm \upmu T}$ directed at angle $\theta=\phi=45\degree$. Cones along $\hat{z}$ are not shown but are present in the simulation. (b) ODMR spectra obtained for $|\vec{B}_{\rm applied}|=50~{\rm \upmu T}$ at $\phi=60\degree$ as a function of $\theta$. For most field angles, all 24 ODMR peaks are resolved. (c) Simulated ODMR spectra under the experimental conditions in (b). Transition frequencies are computed from Eq.~\eqref{eqn:hamiltonian} using $\vec{B}_{\rm nv}=\epsilon \vec{B}_{\rm applied}$, with $\epsilon=19$.
     }
     \label{fig:simul}
\end{figure}

We first corroborate the isotropic nature of our flux concentrator geometry using finite-element magnetostatic simulation. Figure~\ref{fig:simul}(a) shows a map of the magnetic field under an applied field $|\vec{B}_{\rm applied}|=50~{\rm \upmu T}$, oriented at angles $\theta=45\degree$ and $\phi=45\degree$ as defined in Fig.~\ref{fig:background}(c). Far from the cones, the field is uniform, with the same magnitude and angle as $\vec{B}_{\rm applied}$. In the center of the gap between cones (inset), the field magnitude is enhanced by a factor of $\epsilon\approx19$, while the angle of $\vec{B}_{\rm applied}$ is preserved. Moreover, the enhanced magnetic field is fairly uniform throughout the center of the gap region, indicating the potential for isotropic flux concentration without substantial gradient broadening.

We performed an initial validation of the isotropic flux concentrator design by obtaining ODMR spectra as a function of applied field angle. We set $|\vec{B}_{\rm applied}|=50~{\rm \upmu T}$, as confirmed by the reference fluxgate magnetometer  (\ref{app:calibration_fghh}). We then acquired ODMR spectra for a fixed field azimuthal angle $\phi=60\degree$, varying the polar angle $\theta$. The experimental results are shown in Fig.~\ref{fig:simul}(b). Figure~\ref{fig:simul}(c) shows simulated ODMR spectra under the same experimental conditions. The ODMR transition frequencies are computed from Eq.~\eqref{eqn:hamiltonian} using $\vec{B}_{\rm nv}=\epsilon \vec{B}_{\rm applied}$, with $\epsilon=19$. The simulated spectra are largely in agreement with experiment, indicating no major deviations from isotropic enhancement. Moreover, the experimental ODMR linewidth, $\Gamma\approx1.2~{\rm MHz}$, is within a factor of two of the linewidth in the absence of flux concentrators, confirming the absence of substantial gradient broadening (\ref{app:broadening}).

The experimental spectra in Fig.~\ref{fig:simul}(b) are well resolved, demonstrating relief from spectral congestion. A figure of merit for spectral congestion is the ratio $\gamma_{\rm nv}|\vec{B}_{\rm nv}|/\Gamma$, where a larger value is desirable. In the absence of flux concentrators, taking $\Gamma\,{\approx}\,1.2~{\rm MHz}$ at $|\vec{B}_{\rm applied}|\,{=}\,50~{\rm \upmu T}$, this ratio is $\gamma_{\rm nv}|\vec{B}_{\rm applied}|/\Gamma\,{\approx}\,1.2$, which is insufficient to resolve ODMR peaks and accurately determine $\vec{B}_{\rm applied}$. In our experiments, with $\epsilon=19$, we realize a ratio $\gamma_{\rm nv}\epsilon |\vec{B}_{\rm applied}|/\Gamma\,{\approx}\,22$, which is sufficient to resolve all 24 ODMR peaks for most field angles.

\section{Angular dependence and accuracy}
\label{sec:angular}
We next characterize the angular dependence of the device in detail. A magnetic field $|\vec{B}_{\rm fg}(\theta,\phi)|\,{=}\,50~{\rm \upmu T}$ is applied. Here, $\vec{B}_{\rm fg}(\theta,\phi)$ is the field vector recorded by the reference fluxgate magnetometer after accounting for a small (${\sim}1~{\rm \upmu T}$) persistent offset field between the fluxgate and diamond magnetometer positions (\ref{app:offset}). We record an ODMR spectrum ($6~{\rm s}$ of signal averaging) and fit it to determine the field within the NV illumination region, $\vec{B}_{\rm nv}$, making use of separate measurements of the orientation of the NV axes with respect to the lab frame (\ref{app:offset}). We then change $(\theta,\phi)$ in $10\degree$ increments and repeat until the full $4\pi$ solid angle has been mapped twice.

\begin{figure*}[hbt]
    \includegraphics[width=0.99\textwidth]{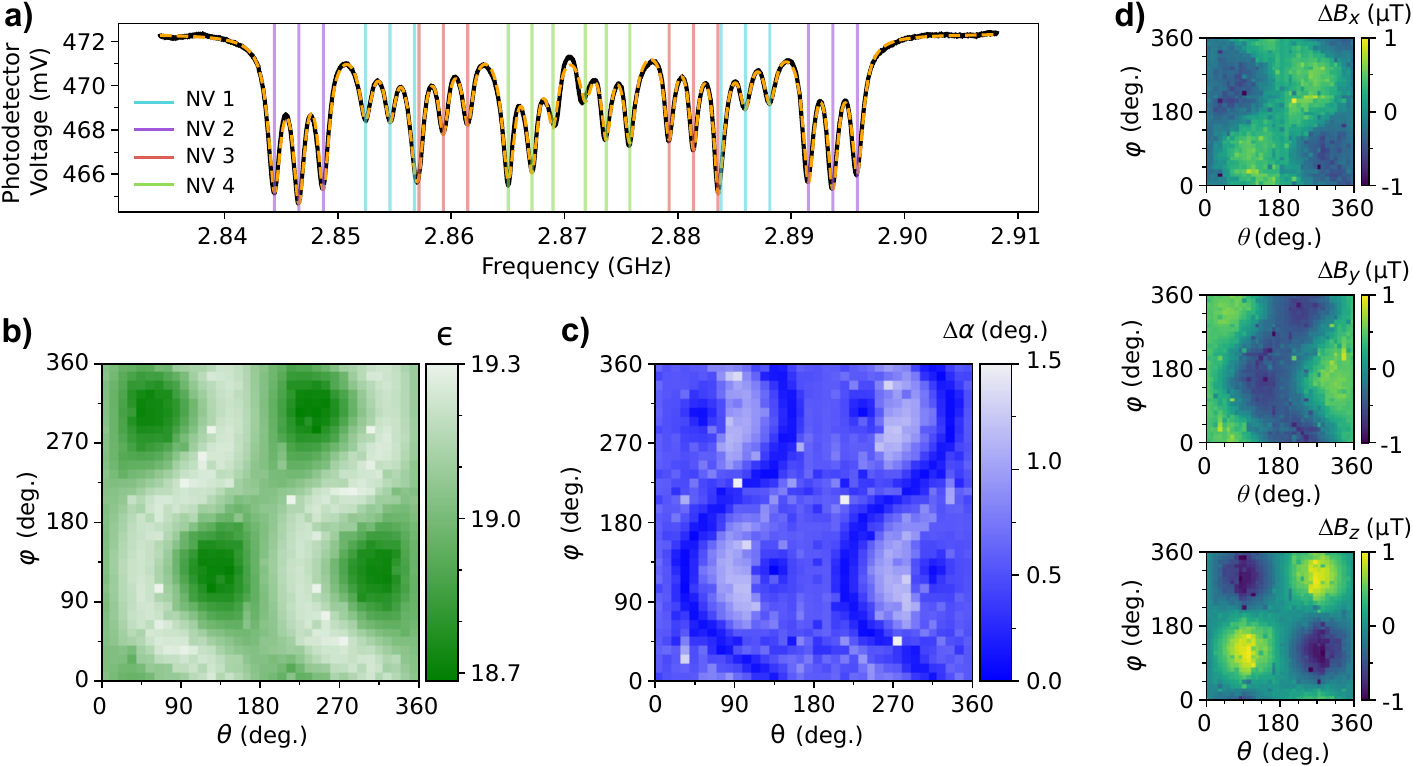}
     \caption{\textbf{Magnetometer angular response.} (a) Example experimental ODMR spectrum (black) and fit (dashed orange). The ODMR central frequencies of each NV axis are annotated with colored vertical bars. (b) Angular dependence of the enhancement factor, $\epsilon(\theta,\phi)$. After weighting measurements by their solid angle coverage, the mean enhancement is $19.05$ and standard deviation is $0.16$.  (c) Angular dependence of the field angle error, $\Delta\alpha(\theta,\phi)$. The weighted mean error is $\bar{\Delta \alpha}=0.57\degree$. (d) Field vector errors, $\Delta \vec{B}=\vec{B}_{\rm nv}/\bar{\epsilon}-\vec{B}_{\rm fg}$, under an assumption of an isotropic concentrator with $\bar{\epsilon}=19.05$. Errors in the estimated field vector can be reduced using a pre-calibrated matrix representation of the enhancement to describe residual anisotropy (\ref{app:fit_enhtensor}).}
    \label{fig:results}
\end{figure*}

Figure~\ref{fig:results}(a) shows an example ODMR spectrum along with its fit. The fit function comprises 24 Lorentzian peaks with variable contrast, $C$, and FWHM linewidth, $\Gamma$. The Lorentzian central frequencies are constrained to be a set of allowed spin transition frequencies obtained from the Hamiltonian in Eq.~\eqref{eqn:hamiltonian}, see~\ref{app:fit_hamiltonian}. All parameters in Eq.~\eqref{eqn:hamiltonian} are fixed based on either literature values ($A_{\parallel}$, $P$, $\gamma_{\rm nv}$, $\gamma_{\rm nuc}$) or independent measurements ($E$). The only exception is the axial zero-field splitting parameter, $D$, which is allowed to vary within $2870.08\pm0.04~{\rm MHz}$ to account for its dependence on ambient temperature variation. Owing to the optical and microwave polarization dependence of the NV spin transitions, each NV alignment has a distribution of $C$ and $\Gamma$ over $\{\theta,\phi\}$ that is somewhat distinct from the others (\ref{app:fit_statistics}). These distributions are statistically deduced and used to aid the fit initial parameters and constraints.  

From the ODMR spectrum fits at each applied field angle, we compute several metrics to characterize the accuracy of the device. The quantity $\epsilon=|\vec{B}_{\rm nv}|/|\vec{B}_{\rm fg}|$ is the scalar enhancement factor. For a perfectly isotropic flux concentrator, $\epsilon$ would be independent of field angle. The experimental $\epsilon(\theta,\phi)$ is plotted in Fig.~\ref{fig:results}(b). While there is some variation of $\epsilon$ with field angle, it is relatively small, indicating nearly isotropic behavior. When weighting each measurement by its solid-angle coverage (\ref{app:solidangle}), we find a mean value $\bar{\epsilon}=19.05$ and standard deviation $\sigma({\epsilon})=0.16$, corresponding to a fractional variation $\sigma({\epsilon})/\bar{\epsilon}=8\times10^{-3}$. 

Another metric is the field angle error, defined as $\Delta \alpha=\cos^{-1}[\vec{B}_{\rm fg}\cdot \vec{B}_{\rm nv}/(|\vec{B}_{\rm nv}||\vec{B}_{\rm fg}|)]$. This metric captures undesired rotations of the field angle due to flux concentrator anisotropy. Figure~\ref{fig:results}(c) plots $\Delta \alpha(\theta,\phi)$. When weighting for solid-angle coverage, we find a mean error of $\bar{\Delta \alpha}=0.57\degree$.

Next, we compute errors for each of the field vector components, given by $\Delta \vec{B}=\vec{B}_{\rm nv}/\bar{\epsilon}-\vec{B}_{\rm fg}$, under the assumption of isotropic enhancement, $\bar{\epsilon}=19.05$. The results for each field component are shown in Fig.~\ref{fig:results}(d). While the deviations are relatively small ($\lesssim1~{\rm \upmu T}$ for a $50~{\rm \upmu T}$ applied field), a systematic fixed pattern is present. Variations in $\Delta \vec{B}(\theta,\phi)$ correspond to weighted standard deviations of $\{0.34,0.38,0.49\}~{\rm \upmu T}$ for $\Delta B_{x,y,z}$ respectively. 

While these metrics capture deviations from isotropicity, they do not represent fundamental limits on the magnetometer accuracy. In \ref{app:fit_enhtensor}, we show that the residual angular dependence of $\epsilon$ can largely be accounted for by representing the magnetic field within the NV illumination region as $\vec{B}_{\rm nv}=\mathcal{E}\vec{B}_{\rm fg}$, where $\mathcal{E}$ is a $3\times3$ transformation matrix. Moreover, $\mathcal{E}$ can be accurately estimated using as few as three measurements of $\vec{B}_{\rm nv}$ obtained under fields applied in different directions. A six-measurement calibration using orthogonal applied fields provides slightly better accuracy. The resulting diamond magnetometer field estimate is $\mathcal{E}^{-1}\vec{B}_{\rm nv}$, and field vector errors are given as $\Delta \vec{B}=\mathcal{E}^{-1}\vec{B}_{\rm nv}-\vec{B}_{\rm fg}$. With the six-measurement calibration, we find that variations in $\Delta \vec{B}(\theta,\phi)$ correspond to weighted standard deviations of $\{0.14,0.11,0.13\}~{\rm \upmu T}$ for $\Delta B_{x,y,z}$ respectively. The weighted mean of angle errors becomes $0.19\degree$ and the weighted standard deviation of the relative field magnitude errors is $\sigma(|\Delta \vec{B}|/|\vec{B}_{\rm fg}|)=2.6\times10^{-3}$. Each corresponds to a factor of ${\sim}3$ reduction compared to the use of a scalar enhancement factor.

\section{Estimation theory analysis of deadzones}
\label{sec:crb}
For some applied field angles, the ODMR spectra are difficult to fit. While using knowledge of neighboring fit results to improve initial parameters (\ref{app:fit_statistics}) enables the fits to converge, the resulting fit parameters appear as discontinuities in Fig.~\ref{fig:results}(b-d). We noticed that these field angles correspond to ODMR spectra with significant spectral overlap. We selected a small region of the solid angle where such overlap occurs for further investigation.

\begin{figure*}[hbt]
    \includegraphics[width=.99\textwidth]{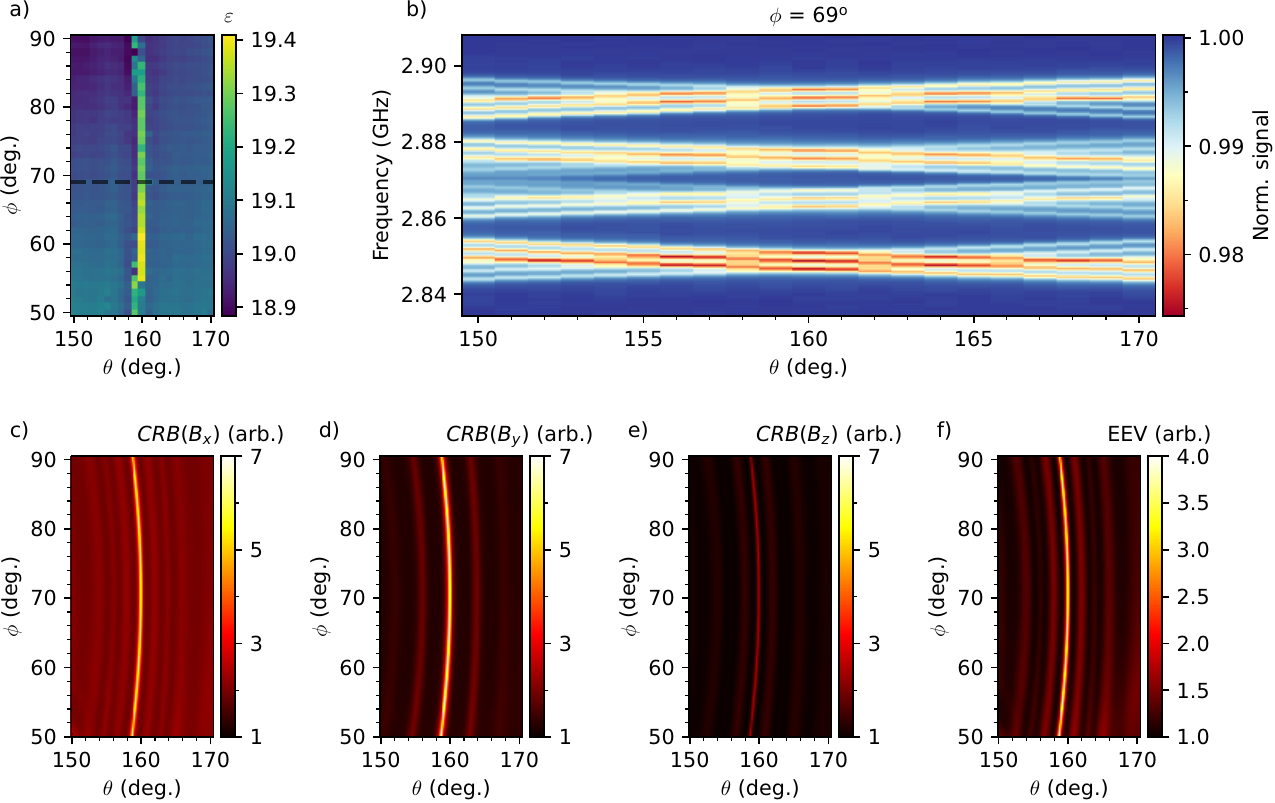}
    \caption{\textbf{Deadzone analysis.} (a) Fitted value of $\epsilon(\theta,\phi)$ over a small angular region with significant spectral overlap. A sharp discontinuity in fit results is observed for $\theta\approx160\degree$. (b) ODMR spectra along the linecut in angular space indicated in part (a). At the edges of the sweep, all 24 ODMR lines are resolved, but at $\theta\approx160\degree$ only 12 peaks can be discerned. (c) Cram{\`e}r-Rao lower bound for estimation uncertainty of $B_x$, calculated for $|\vec{B}_{\rm nv}|=1~{\rm mT}$ over the same angular region as (a). (d,e) Same as (c) except for $B_y$ and $B_z$, respectively. (f) Error ellipsoid volume, EEV, over the same angular region as (a,d-e).
}
    \label{fig:deadzone}
\end{figure*}

Figure~\ref{fig:deadzone}(a) shows the fitted value of $\epsilon$ over this region, with fine increments in $(\theta,\phi)$ of $1\degree$. Sharp discontinuities in the fitted value of $\epsilon$ appear at $\theta\approx160\degree$. Figure~\ref{fig:deadzone}(b) shows the ODMR spectra as a function of $\theta$ for fixed $\phi=69\degree$. It is seen that the ODMR peaks of two pairs of NV axes overlap at $\theta\approx160\degree$, and the number of resolvable ODMR peaks is reduced from 24 to 12. These data suggest a strong correlation between fit error and spectral overlap, which we confirm by analyzing fit statistics in \ref{app:fisher}.

To gain further insight into the angular dependence of the magnetometer and the undesired impact of spectral overlap, we apply a Cram{\`e}r-Rao bound analysis~\cite{IRE2005}. As shown in \ref{app:fisher}, a $3\times3$ Fisher information matrix can be expressed as:
\begin{equation}
  \mathcal{I}_{p,q}(\vec{B}_{\rm nv})\propto\sum_{i=1}^{i_{\rm max}}\frac{\partial F_{f_i}(\vec{B}_{\rm nv})}{\partial B_p}\frac{\partial F_{f_i}(\vec{B}_{\rm nv})}{\partial B_q},
   \label{eq:fishermain} 
\end{equation}
where  $F_{f_i}(\vec{B}_{\rm nv})$ is the ODMR spectral response for microwave frequency $f_i$ in a magnetic field $\vec{B}_{\rm nv}$, $i_{\rm max}$ is the number of points in the spectrum, and $p$ and $q$ are indices spanning the $\{x,y,z\}$ Cartesian axes. Expression~\eqref{eq:fishermain} is derived assuming the noise at every point in the ODMR spectrum has the same Gaussian or Poissonian distribution. For a given field within the NV illumination region, $\vec{B}_{\rm nv}$, the Cram{\`e}r-Rao lower bound on estimation uncertainty of vector component $B_p$ is calculated from the diagonal elements of the inverse Fisher information matrix as $\mathrm{CRB}(B_p)=\sqrt{[\mathcal{I}^{-1}]_{pp}}$. 

We computed the relative Cram{\`e}r-Rao bounds for each field component at all possible angles of $\vec{B}_{\rm nv}$, taking $|\vec{B}_{\rm nv}|=1~{\rm mT}$. The results are shown in Fig.~\ref{fig:deadzone}(c-e) for the same angular sub-region as explored experimentally in Fig.~\ref{fig:deadzone}(a). A peak in the Cram{\`e}r-Rao bounds of each field component is observed at $\theta\approx160\degree$, indicating that the theoretical uncertainty in field estimation is greatest for these field angles. We also compute an additional metric, the error ellipsoid volume~\cite{HUT2001} that captures a combined uncertainty across all three field components, ${\rm EEV}=1/\sqrt{{\rm det}(\mathcal{I})}$. The EEV of the same sub-region is plotted in Fig.~\ref{fig:deadzone}(f). It also exhibits local maxima at $\theta\approx160\degree$.

In \ref{app:fisher}, we show plots of the calculated $\mathrm{CRB}(B_{x,y,z})$ for all angles $(\theta,\phi)$, taking $|\vec{B}_{\rm nv}|=1~{\rm mT}$. The theoretical bounds and the experimental fit uncertainties exhibit peaks at the same angular positions, corresponding to zones of substantial spectral overlap. Notably, the exact $\mathrm{CRB}$ values in these zones depend on assumptions of the ODMR contrast. The calculations presented in Fig.~\ref{fig:deadzone} and \ref{app:fisher} assume each NV axis has a distinct contrast, due to the polarization anisotropy of the NV optical and microwave transitions. Specifically we choose the average contrast for a given NV axis observed in the experiments of Fig.~\ref{fig:results}. In this case, the $\mathrm{CRB}(B_{x,y,z})$ values in regions of spectral overlap are significantly smaller than those when we assume every NV axis has the same contrast. This indicates that the polarization anisotropy of the NV optical and microwave transitions is a valuable resource in minimizing deadzones.

A metric for quantifying the boundaries of angular ``deadzones'' in a magnetometer can be constructed by evaluating where $\mathrm{CRB}(B_{x,y,z})$ is a factor of $\geq2$ of those of the fully-resolved case. From the analysis in \ref{app:fisher}, we find that this deadzone condition holds for $1.7\%$ of the total $4\pi$ solid angle for estimation of $B_x$, $2.1\%$ for $B_y$, and $2.3\%$ for $B_z$. While a sufficiently large bias field can result in a more flat angular dependence~\cite{SCH2018,SCH2025}, the deadzone fraction of our device is comparable to or better than our estimates for other zero-bias vector magnetometers based on spin resonance reported in the literature (\ref{app:deadzonescrape}), such as alkali-metal vapor sensors ($1-35\%$~\cite{BEN2010}) and Overhauser nuclear magnetic resonance magnetometers ($10-50\%$~\cite{GE2020}). The deadzone fraction could be further reduced by using a smaller NV illumination volume to reduce gradient broadening and modifying the concentrator geometry to provide a larger enhancement factor. 

\section{Temporal stability}
\label{sec:temporal}
\begin{figure}[htb]
  \includegraphics[width=\columnwidth]{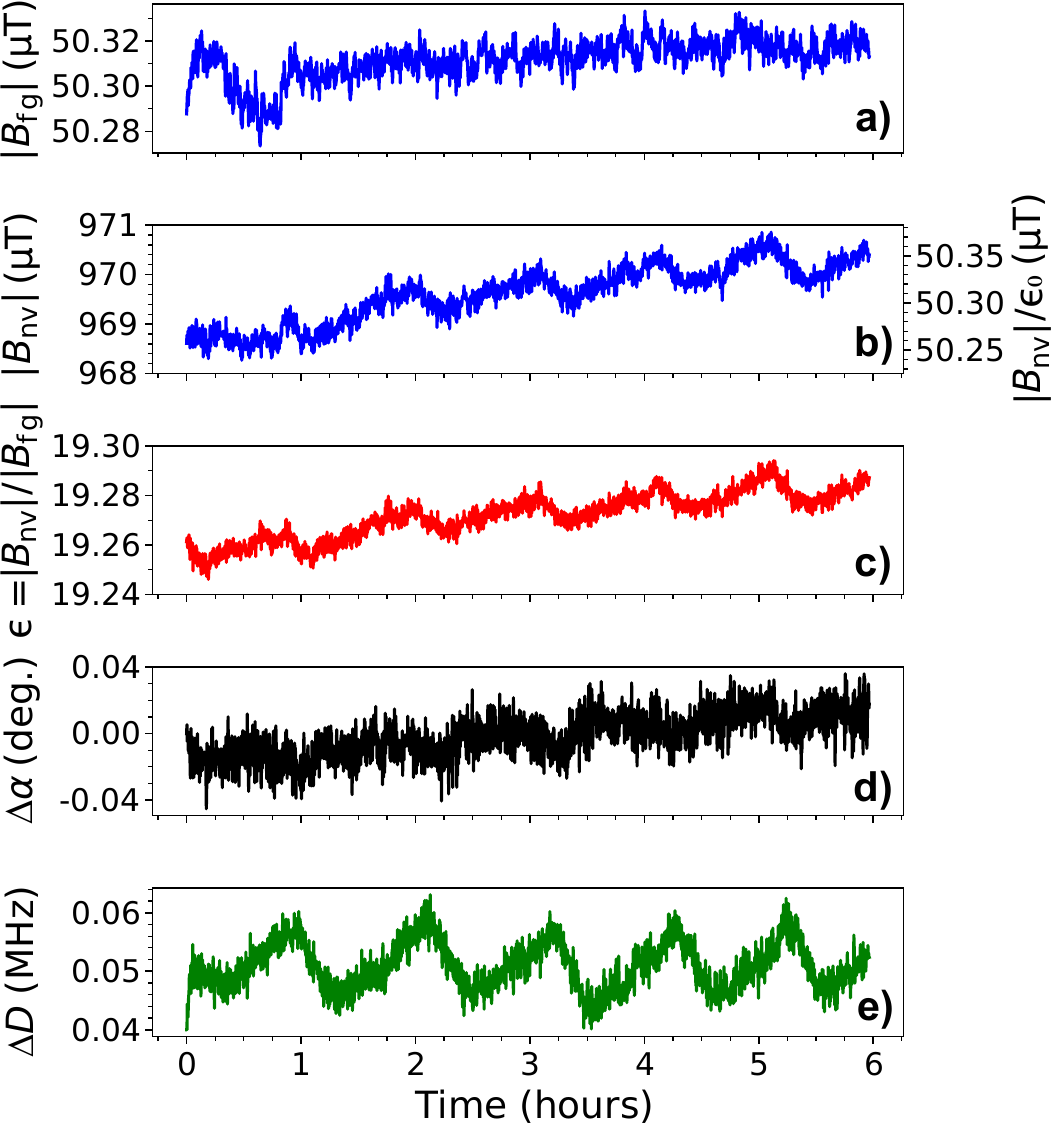}
  \caption{\textbf{Sensor stability.}  (a) Fluxgate magnetometer field magnitude, $|\vec{B}_{\rm fg}|$, versus time.  (b) Time dependence of $|\vec{B}_{\rm nv}|$, as measured by ODMR spectroscopy ($6~{\rm s}$ of signal averaging for each spectrum).  The right axis represents a field estimate, assuming isotropic enhancement factor $\epsilon_0\,{=}\,19.27$.  (c) Scalar enhancement factor, $\epsilon$, versus time. (d) Field angle error, $\Delta \alpha\,{=}\,\cos^{-1}(\vec{B}_{fg}\cdot \vec{B}_{nv})/(|\vec{B}_{nv}||\vec{B}_{fg}|$, versus time. (e) Time dependence of the fitted zero-field splitting, where $\Delta D=D-2870~{\rm MHz}$.
  }
  \label{fig:stability}
\end{figure}

Finally, we tested the stability of the magnetometer by recording ODMR spectra continuously over a period of six hours. The field is initially set to $[35.18,-30.43,-16.83]~{\rm \upmu T}$, resulting in a well-resolved spectrum. Figure~\ref{fig:stability}(a) shows the magnitude of the fluxgate reference magnetometer signal as a function of time. The field was largely stable to $\lesssim40~{\rm nT}$ over the course of the measurement. Figure~\ref{fig:stability}(b) shows the fitted field magnitude within the diamond illumination region, $|\vec{B}_{\rm nv}|$ as a function of time. Here, some additional drift is observed as well as an oscillatory pattern with a period of ${\sim}1~{\rm hour}$. We convert this signal to an external field magnitude estimate by dividing by a scalar enhancement factor, $\epsilon_0=19.27$, that is appropriate for this field angle. The variations in field estimate $|\vec{B}_{\rm nv}|/\epsilon_0$ are ${\sim}100~{\rm nT}$ over the six hour measurement, somewhat larger than that of the fluxgate reference. This corresponds to a variation of enhancement factor, $\epsilon=|\vec{B}_{\rm nv}|/|\vec{B}_{\rm fg}|$, over a range of approximately $19.25{\mbox{-}}19.29$, Fig.~\ref{fig:stability}(c). Nevertheless, the variations in field magnitude estimate do not exceed ${\sim}40~{\rm nT}$ over any given hour, which is comparable to or better than prior diamond magnetometer works~\cite{FES2020,ZHA2021,GAO2023,GRA2025}. Moreover, we track the stability of the field angle error $\Delta \alpha$, Fig.~\ref{fig:stability}(d), and find a drift of $\lesssim0.04\degree$ over the course of six hours, indicating promise for absolute diamond vector magnetometry~\cite{FAN2013,SCH2018,LON2025}.

We attribute the ${\sim}1~{\rm hour}$ period oscillation present in Figs.~\ref{fig:stability}(b,c) to variation in the ambient temperature. Figure~\ref{fig:stability}(e) shows the time dependence of the zero field splitting. A clear oscillation is observed with a peak-to-peak amplitude of ${\sim}10~{\rm kHz}$. Given the known temperature dependence of $D$~\cite{ACO2010}, this corresponds to a temperature oscillation with a peak-to-peak amplitude of ${\sim}0.14~{\rm K}$. The oscillation period and amplitude are consistent with the lab temperature feedback control, which was characterized by monitoring $D$ and the lab temperature versus time on different nights (\ref{app:temperature}). Since variations in $D$ should not affect the magnetic field estimation, we hypothesize that the oscillations present in Figs.~\ref{fig:stability}(b,c) are due to thermal expansion of the flux concentrators~\cite{FES2020}. This is a potential source of systematic error that might be mitigated in the future by calibration or mechanical feedback.

\section{\label{sec:discussion}Discussion}
Our study highlights the promise of isotropic flux concentrators to reduce spectral congestion in magnetic resonance and quantum sensing applications. We demonstrated a device that can accurately detect Earth's magnetic field with magnitude errors at the ${\sim}100~{\rm nT}$ level and angle errors $\lesssim0.2\degree$. This is already comparable to some commercial triaxial magnetoresistance sensors~\cite{VOP2003,WAN2024,ANA2014,BOS2025} and approaches the offset and orthogonality errors of some commercial fluxgate magnetometers~\cite{HU2024,CHO2024}. In contrast to those triaxial magnetometers, our device offers the advantage of simultaneously measuring all magnetic field vector components at a single position. Moreover, the magnetic sensitivity is expected to be ${\sim}19$-times better than a diamond vector magnetometer without flux concentrators using the same optical power~\cite{FES2020}.

Several challenges still remain, and overcoming them will be important for field applications. First, the concentrator demonstrated here is not perfectly isotropic, as the angle-dependence of the enhancement factor has a relative standard deviation of $0.8\%$. While pre-calibration improves the accuracy, a more mechanically precise device (\ref{app:simulation}) would be advantageous. Second, the temperature dependence of the concentrator leads to drifts in field estimation at the level of ${\sim}40~{\rm nT/hour}$ in our experiments. This could be improved by better temperature control or data correction. Finally, our study performed ODMR spectroscopy with a relatively broad frequency span ($80~{\rm MHz}$) and slow repetition rate ($5~{\rm Hz}$), see \ref{app:save_setup}. The measurements were not optimized for high sensitivity. Instead, we used signal averaging to ensure that the sensitivity was adequate to resolve absolute accuracy errors. While this is useful for characterizing the device accuracy and angular dependence, a practical field magnetometer could benefit from using multifrequency microwave modulation methods~\cite{SCH2018} for sensitive field vector estimation. The use of higher laser power and higher numerical aperture collection optics would also improve the magnetic sensitivity, potentially reaching the ${\sim}1~{\rm pT\,s^{1/2}}$ level~\cite{FES2020}.

In summary, we have demonstrated how a nearly isotropic flux concentrator can be used to perform diamond vector magnetometry at Earth's magnetic field. We characterized the angular dependence of the device and identified critical factors for further improvement. We anticipate our method may find application in geomagnetic surveys, magnetic anomaly detection, and inertial navigation systems.

\begin{acknowledgments}
We gratefully acknowledge advice and support from P.~Kehayias, I.~Savukov, M.~Malone, N. Solmeyer, L. Zipp, D.~Lidsky, and A.~Jeronimo~Perez. We especially thank A.~Gravagne for helping to design and machine the PEEK holder.\\
\textbf{Competing interests.} The authors declare no competing financial interests.\\
\textbf{Author contributions.} I.F., M.S.Z., N.M., and V.M.A. conceived the idea and designed the experiments. M.S.Z. and I.F. assembled the cones and diamond apparatus. N.M. and M.S.Z. built the ODMR apparatus and acquired data.  M.S.Z. wrote the control and analysis software and analyzed the data. I.F., N.M., Y.S., and M.S.Z. conducted magnetostatic modeling. B.A.R., A.B., M.A., K.A.L., A.J., and J.S. assisted in theoretical modeling, experimental execution, and data interpretation. M.S.Z. and V.M.A wrote the initial manuscript draft. V.M.A. supervised the project. All authors helped edit the manuscript. \\
\textbf{Funding.} This work was supported by the Moore Foundation (EPI-12968), National Science Foundation (CHE-1945148), Department of Energy (LDRD-20220086DR), and National Institutes of Health (DP2GM140921, R41EB036583). I.F. acknowledges support from ERDF project 1.1.1.3/1/24/A/166.

\end{acknowledgments}

\clearpage
\appendix
\setcounter{equation}{0}
\setcounter{section}{0}
\makeatletter
\renewcommand{\thetable}{A\arabic{table}}
\renewcommand{\theequation}{A\Roman{section}-\arabic{equation}}
\renewcommand{\thefigure}{A\arabic{figure}}
\renewcommand{\thesection}{Appendix~\Roman{section}}
\makeatother

\section{\label{app:setup}Experimental Setup}

\subsection{\label{app:save_setup}Tabletop apparatus}

The experimental setup is shown in Fig.~\ref{fig:background} of the main text.  Here we provide additional details. The polarization of a 532-nm diode-pumped solid-state laser (Thorlabs DJ532-40) beam is adjusted via waveplates.  The laser beam reflects off of a dichroic mirror (Thorlabs DMLP567L) and is focused by a one-inch-diameter NA$=0.3$ lens. The converging laser light passes through a ${\sim}25~{\rm mm}$ diameter hole in the corner facet of the cubic PEEK housing, then through a $500\mbox{-}{\rm \upmu m}$-diameter sapphire halfball lens (Edmund Optics, 49-553), and finally is incident on the diamond membrane. We estimate that ${\sim}10~{\rm mW}$ of laser light is incident on the diamond. The laser spot size in the diamond is ${\sim}50~{\rm \upmu m}$ diameter. The laser spot area, together with the ${\sim}50~{\rm \upmu m}$ thickness of the diamond membrane, define the NV illumination region.  

Red NV fluorescence from the diamond passes through the halfball lens and is collimated by the NA$=0.3$ lens. The fluorescence beam then passes through the dichroic mirror and is filtered through a $650~{\rm nm}$ long-pass filter (Thorlabs FELH0650). Finally, the fluorescence light passes through an adjustable linear polarizer (Thorlabs LPVISE100-A) and is focused by a NA$=0.78$ lens (Thorlabs ACL12708) onto an amplified photodetector with variable gain (Newport New Focus 2051). The photodetector output is connected to an oscilloscope (Yokogawa DL9040L) and ODMR sweeps are triggered by the ramp of the microwave signal generator.

\begin{figure}[hbt]
    \includegraphics[width=\columnwidth]{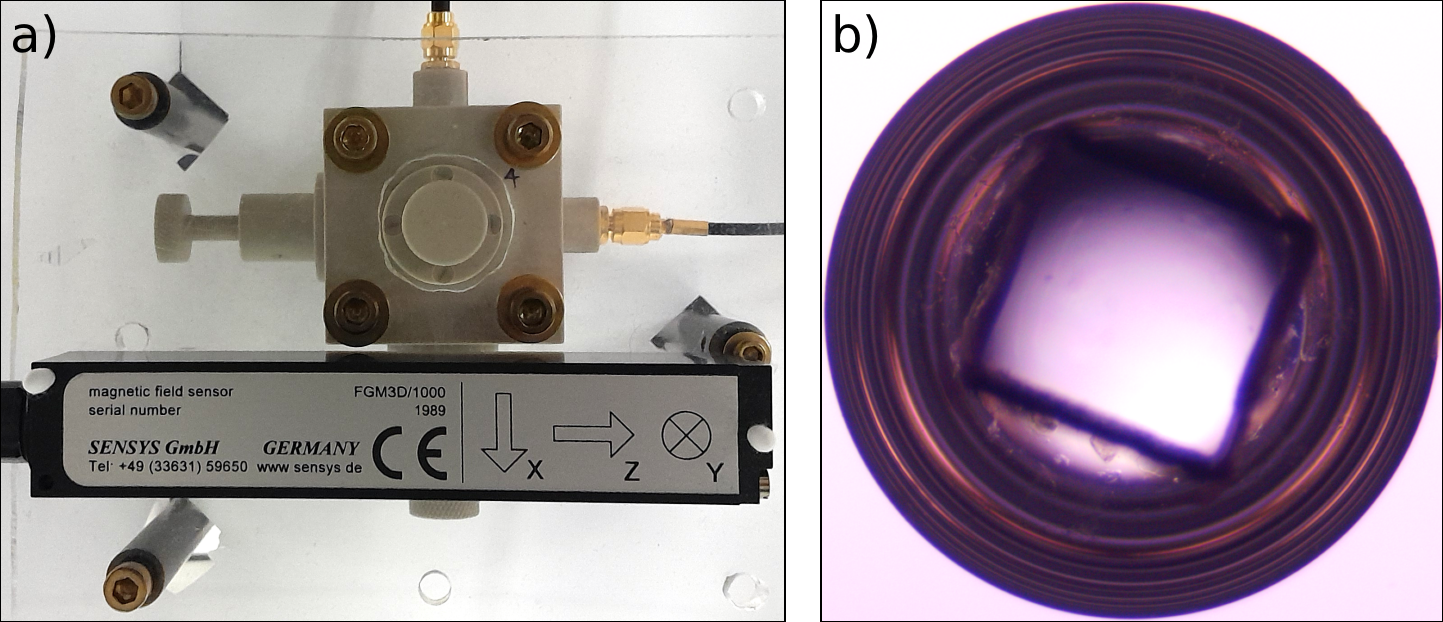}
    \caption{\textbf{Photographs of the assembly.} (a) PEEK housing (cubic, ${\sim}2~{\rm inch}$ outer dimensions) and triaxial fluxgate magnetometer, as placed during experiments.  The heads of the fine-thread screws used to translate the adjustable cones are visible.  (b) Diamond membrane affixed to the $500\mbox{-}{\rm \upmu m}$-diameter sapphire halfball lens.}
    \label{fig:assembly}
\end{figure}

A triaxial fluxgate magnetometer (Sensys FGM3D 1000) provides reference field measurements.  The fluxgate magnetometer and PEEK housing are attached to a plexiglass sheet. A photograph of the PEEK housing and fluxgate magnetometer is shown in Fig.~\ref{fig:assembly}(a).  

The diamond membrane is cut from a $^{12}$C-enriched diamond, grown by chemical vapor deposition, that was doped with ${\sim}4~{\rm ppm}$ NV centers via electron irradiation and annealing. The large faces of the membrane are polished normal to a [110] crystallographic direction.  Similar pieces of this diamond were used in Ref.~\cite{SMI2025}. The [110] surface-normal directions point approximately parallel to the $\hat{x}+\hat{y}+\hat{z}$ direction of the lab frame.  The diamond membrane (dimensions: ${\sim}250\times250\times50~{\rm \upmu m}$) is centered on the flat surface of the halfball lens and fixed using UV-curing epoxy. A photograph of the diamond affixed to the halfball lens is shown in Fig.~\ref{fig:assembly}~(b).

Enameled copper wire (38-gauge) is used to deliver microwaves. Three separate wires are wound around the three fixed ferrite cones' tips. The wires are soldered to SMA connectors affixed to the PEEK housing. 

By varying the laser light polarization, fluorescence polarization, and microwave polarization, we are able to vary the relative ODMR contrast of each NV axis. We select a combination of polarizations that allows for appreciable ($>0.5\%$) contrast for every ODMR line, see Fig.~\ref{fig:results}(a) of the main text. This choice ensures that the magnetic sensitivity of each field component is similar for most applied field angles.

A set of three-axis Helmholtz coils (Serviciencia, Ferronato BH600-3-B) with ${\sim}0.6~{\rm m}$ coil diameters encloses the diamond and cone assembly. Varying the current in the coils allows us to simulate Earth's magnetic field with full vector control.

All ODMR spectra analyzed in the main text, Figs.~\ref{fig:simul}(b), \ref{fig:results}, \ref{fig:deadzone}(a,b), and \ref{fig:stability}, used the same microwave sweeping sequence. The signal frequency from a signal generator (Stanford Research Systems SG384) is repeatedly swept from $2.83$ to $2.91~{\rm GHz}$ in a ramp waveform, at a rate of $4.86~{\rm Hz}$. The microwave power output by the signal generator ($-9~{\rm dBm}$) is amplified by a power amplifier (Mini-Circuits ZHL-16W-43-S+). The signal then passes through a $2\mbox{-}4~{\rm GHz}$ circulator (Pasternack PE8401) and is fed to the copper wires wrapped around the fixed cone tips.

\subsection{\label{app:assembly}Flux concentrator design and construction}

The design of the isotropic flux concentrator involves optimizing a tradeoff between sensor size, enhancement factor, gradient broadening, and mechanical alignment tolerance. For example, a small sensor is desirable for many field applications. However larger concentrators may be easier to mechanically align, and they feature larger enhancement factors, which (all else equal) improves sensitivity and alleviates spectral congestion. In principle one could make the concentrator small and still realize a large enhancement factor, by shrinking the gap between cones. However, for a finite size NV illumination volume (in our case, approximately ($50~{\rm \upmu m})^3$), this leads to broadening of the ODMR lines (at non-zero applied field) due to spatial variation in the enhancement factor. Once gradient broadening dominates the ODMR linewidth, further increase of the enhancement factor will not improve the sensitivity or alleviate spectral congestion. It may also be harder to align and keep mechanically stable such a configuration.

We conducted a series of parameter sweeps using finite-element magnetostatic modeling to decide on the design. We constrained the entire concentrator device to fit within a ${\sim}30~{\rm mm}$ cube and chose MN60 ferrite as the flux concentrator material~\cite{FES2020}. We defined the NV illumination region as a ${\sim}(50~{\rm \upmu m})^3$ volume in the center of the device. We required that, under a $50~{\rm \upmu T}$ applied field, the standard deviation of the magnetic field within the NV illumination region was ${\lesssim}\,10~{\rm \upmu T}$. This condition corresponds to an estimated gradient broadening contribution to the ODMR FWHM linewidth of less than $1~{\rm MHz}$. We also required that the average enhancement factor within the NV illumination volume did not vary as a function applied field angle by more than $0.1\%$ and the average field angle within the NV illumination volume did not deviate from the applied field angle by more than $0.1\degree$. Finally, we required that the diameter of the ferrite cone tips was never smaller than $0.24~{\rm mm}$, based on expected machining tolerances, and the diameter of the bases was small enough to provide optical access to $>50\%$ of the fluorescence solid angle.

\begin{figure}[hbt]    
\includegraphics[width=0.65\columnwidth]{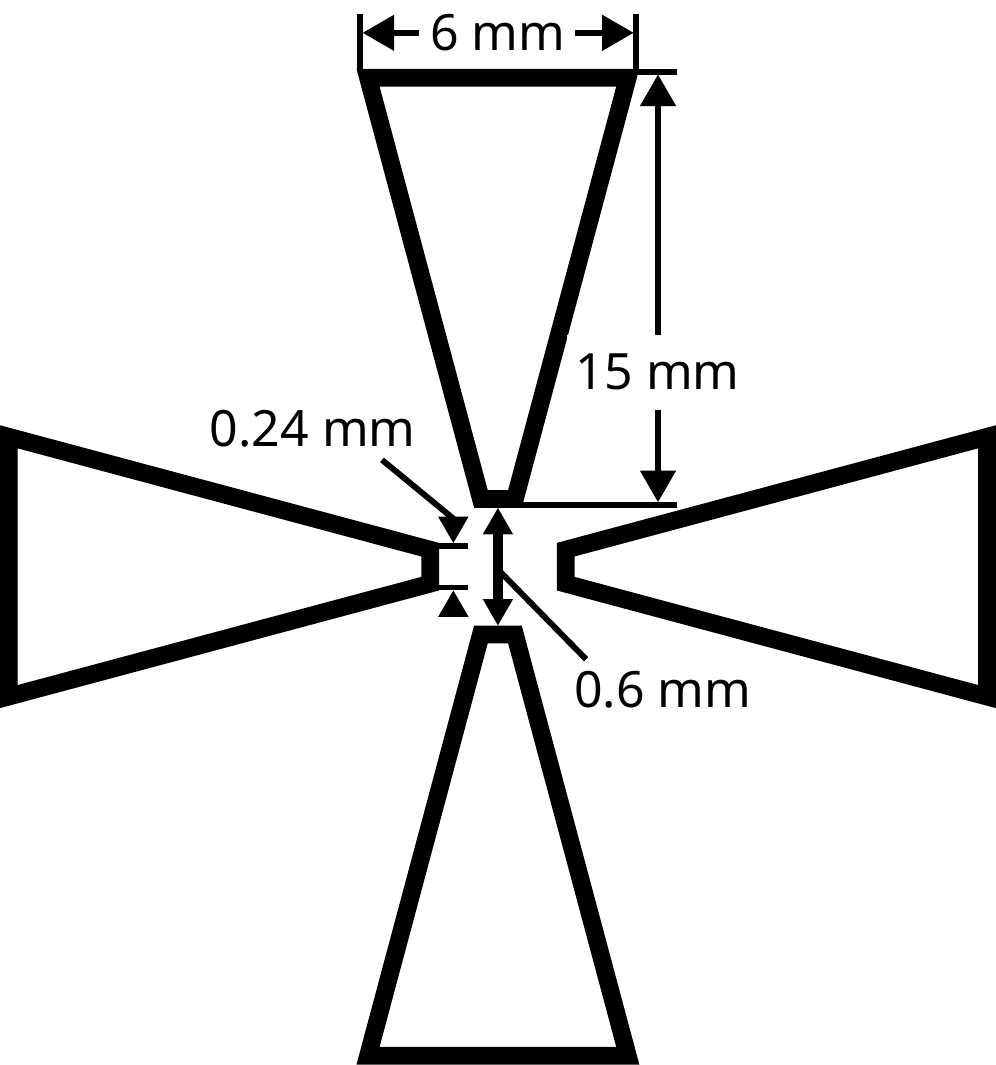}
    \caption{\textbf{Cone dimensions.} Four (out of six) cones are shown for visual clarity. The other pair of cones are aligned normal to the page, with the same dimensions and gaps.}
    \label{fig:conedims}
\end{figure}

\begin{figure}[hbt]
    \includegraphics[width=0.8\columnwidth]{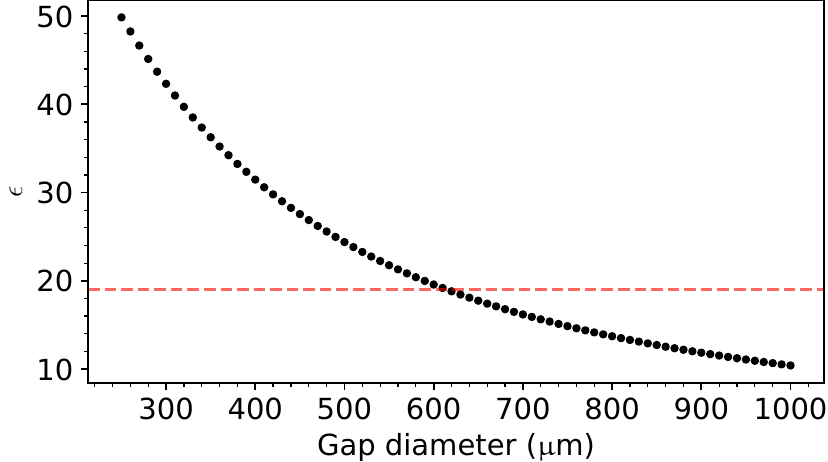}
    \caption{\textbf{Modeling cone gap.} The enhancement factor, $\epsilon$, at the center of the geometry is simulated as a function of the gap between cone tips.
    }
    \label{fig:symmetricgap}
\end{figure}

Under these constraints, the largest simulated enhancement factor is when the cone length is $15~{\rm mm}$. The exact cone tip and base diameters are not very consequential, but a base diameter of $6~{\rm mm}$ and tip diameter of $0.24~{\rm mm}$ is able to satisfy our requirements for gaps ${\gtrsim}\,0.5~{\rm mm}$. The dimensions and spacings of the final design for the ferrite cones are depicted in Fig.~\ref{fig:conedims}. The simulated enhancement factor as a function of gap between cone pairs is shown in Fig.~\ref{fig:symmetricgap}. While the enhancement factor increases monotonically with decreasing gap, gap lengths ${\lesssim}\,0.5~{\rm mm}$ do not meet our gradient broadening requirements and were thus not used. The observed enhancement factor of 19 is consistent with a ${\sim}600~{\rm\upmu m}$ gap between cone tips.

The cones were assembled in the manner as follows. First, the PEEK housing is machined from a $2$-inch cube, and the bases of each of the six cones are affixed to the housing. Next, the tips of the three fixed cones are adhered to the curved surface of the halfball lens. They are configured in such a way that the halfball lens' apex is centered in between the cones to allow for maximum light throughput.

\begin{figure}[hbt]   \includegraphics[width=0.8\columnwidth]{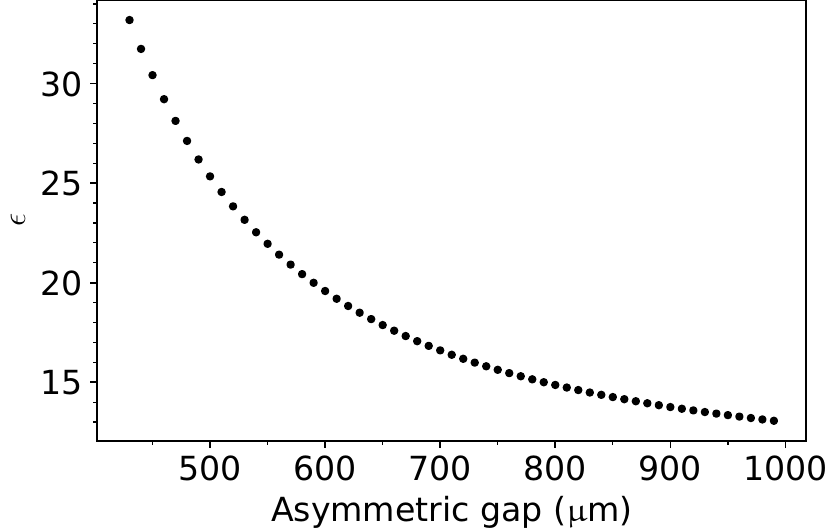}
    \caption{\textbf{Modeling asymmetric gap.} The distance from the center of the NV illumination region to the three fixed cones is fixed at $300~{\rm \upmu m}$. The enhancement factor $\epsilon$ is simulated by simultaneously adjusting the separation of the three adjustable cones with respect to the fixed cones. This separation defines the ``asymmetric gap'' horizontal axis. The desired enhancement factor $\epsilon\approx19$ is realized only when all of the cone tips are ${\sim}300~{\rm \upmu m}$ from the NV illumination region's center. 
    }
    \label{fig:asymmetricgap}
\end{figure}

After the assembly is fabricated, the distance from the center of the NV illumination volume to the fixed cone tips is ${\sim}300~{\rm \upmu m}$. In order to align the adjustable cones, we slowly decrease their gap until the enhancement factor along all axes is ${\sim}19$. Figure~\ref{fig:asymmetricgap} shows the simulated behavior of the enhancement factor as the adjustable cones are translated. As can be seen, the $\epsilon=19$ target is realized only when the adjustable cones are all $300~{\rm \upmu m}$ from the center of the NV illumination region. 

\section{\label{app:broadening}Gradient broadening}
In Sec.~\ref{sec:isotropic} of the main text, we introduced a figure of merit, $\gamma_{\rm nv} |\vec{B}_{\rm nv}/\Gamma|$, to quantify the degree of spectral congestion, where a larger value indicates less spectral congestion. This figure of merit resembles the resolving power of a spectrometer. In our case, the numerator is approximately proportional to the distance between ODMR peaks associated with different NV axes, while the denominator is approximately the spectral spread of each ODMR peak. 

\begin{figure}[hbt]
    \includegraphics[width=\columnwidth]{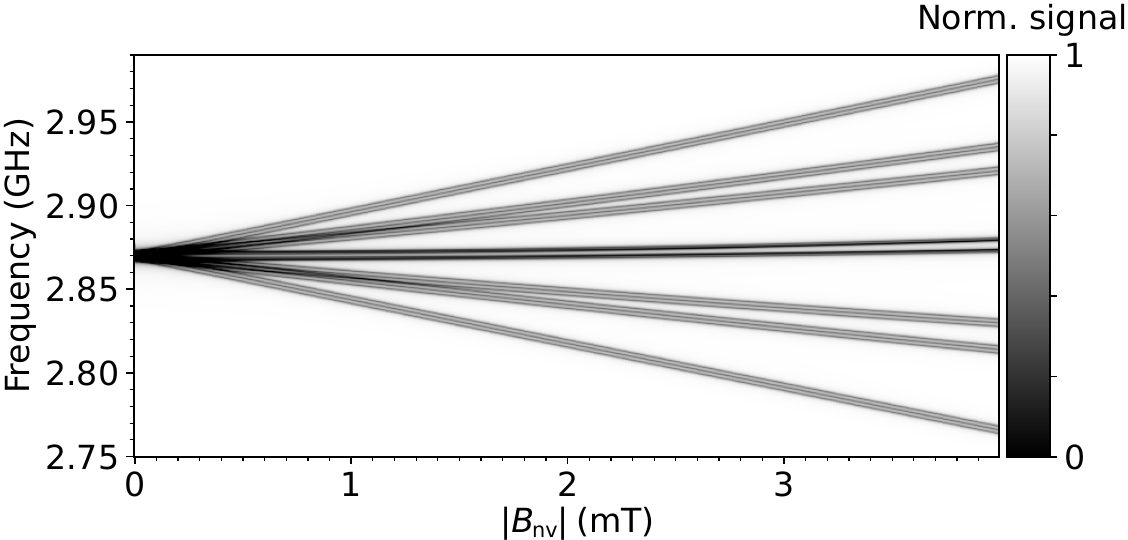}
    \caption{\textbf{Field dependence of ODMR spectra.} Simulation of ODMR spectra as a function of magnetic field magnitude for the same field angle as the data presented in Fig.~\ref{fig:stability} of the main text. 
    }
    \label{fig:lingap}
\end{figure}

Figure~\ref{fig:lingap} shows a plot of the 24 ODMR transition frequencies as a function magnetic field magnitude within the NV illumination volume. Ignoring the triplet hyperfine structure, it is seen that gaps between peaks associated with different NV transition frequencies grow proportional to field magnitude, confirming the logic behind the numerator in the spectral congestion figure of merit. The hyperfine triplet structure means that each transition associated with a particular NV axis is actually comprised of 3 peaks separated by approximately $|A_{\parallel}|=2.17~{\rm MHz}$. This property does not feature in our spectral congestion figure of merit because the hyperfine triplet peak separations are largely independent of field angle and magnitude. In such cases, the triplet can be thought of as just a more complicated lineshape. However a more precise analysis of spectral congestion would include the hyperfine coupling, particularly for characterizing the behavior at very low field.

\begin{figure}[hbt]
    \includegraphics[width=\columnwidth]{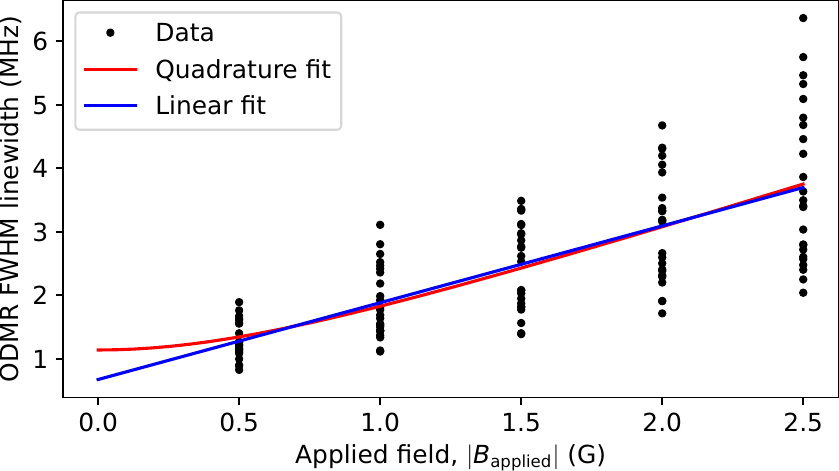}
    \caption{\textbf{ODMR linewidth versus applied field.} The fitted ODMR FWHM, $\Gamma$, of each of the 24 ODMR peaks are plotted as a function of applied field magnitude $|B|$. The data are fit to Expressions~\ref{eqn:allgamma}. The fit to the quadrature-sum model yields $\alpha=1.430\pm0.004~{\rm MHz/G}$ and $\Gamma_0=1.14\pm0.03~{\rm MHz}$, while the fit to the linear model yields $\alpha=1.21\pm0.01~{\rm MHz/G}$ and $\Gamma_0=0.67\pm0.03~{\rm MHz}$. }
    \label{fig:fwhm_b}
\end{figure}

As discussed in \ref{app:assembly}, care was taken to design a flux concentrator that minimized gradient broadening due to spatial variation in enhancement factor. Nevertheless, we observe some broadening to the ODMR lines that scales proportional to field, indicative of spatial variation of enhancement factor within the diamond illumination region. Figure~\ref{fig:fwhm_b} is a plot of the fitted FWHM, $\Gamma$, of each of the 24 ODMR peaks at four applied fields from $50\mbox{-}200~{\rm \upmu T}$ ($0.5\mbox{-}2~{\rm G}$). While there is a substantial spread in the fitted $\Gamma$ for any given field, a clear trend of increasing $\Gamma$ with applied field is observed.

Once $\Gamma$ grows linearly with applied field, the spectral congestion figure of merit saturates and increasing the applied field further does not reduce congestion. It appears that this already happens at or near Earth's field ($50~{\rm \upmu T}$) for our device. The $\Gamma(|B_{\rm applied}|)$ data in Fig.~\ref{fig:fwhm_b} are fit to two models:
\begin{subequations}\label{eqn:allgamma}
\begin{align}
    \Gamma&=\sqrt{(\alpha|B_{\rm applied}|)^2+\Gamma_0^2}\label{eqn:gamma},\\
    \Gamma&=\alpha|B_{\rm applied}|+\Gamma_0\label{eqn:gammalinear},
\end{align}
\end{subequations}
where $\Gamma_0$ is the intrinsic power-broadened FWHM and $\alpha$ is a scale factor. The quadrature-sum model is appropriate if the ODMR lines and field distribution can both be described by Gaussians, while the linear model is appropriate if the lineshapes can both be described by Lorentzians. Either model fits well to the data, with the main difference being the fitted values of the intrinsic linewidth, $\Gamma_0=0.67\pm0.03~{\rm MHz}$ for the quadrature-sum model and $\Gamma_0=1.14\pm0.03~{\rm MHz}$ for the linear model. We expect the true lineshape to be closer to a Voigt profile, in which case the true zero-field FWHM would be somewhere between these values. In any case, it is expected from the data in Fig.~\ref{fig:fwhm_b} that $\Gamma_0$ is within a factor of two of $\Gamma(50~{\rm \upmu T})\approx1.2~{\rm MHz}$, as assumed in estimates of spectral congestion in Sec.~\ref{sec:isotropic} of the main text.

\begin{figure}[hbt]
    \includegraphics[width=\columnwidth]{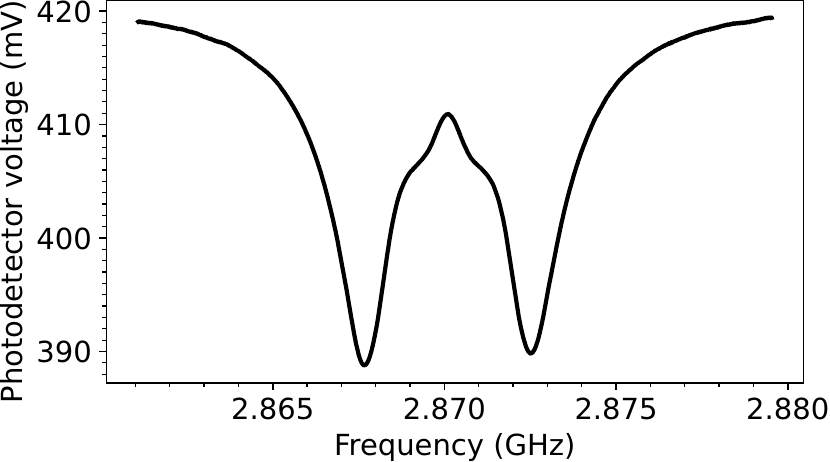}
    \caption{\textbf{Zero-field ODMR spectrum.} The magnetic field was hand-tuned to cancel the ambient magnetic field. The apparent FWHM is ${\sim}1.5~{\rm MHz}$. The splitting between the two peaks is $2|A_{\parallel}|=4.34~{\rm MHz}$. The splitting between the inner shoulders is $2E\approx2~{\rm MHz}$. 
    }
    \label{fig:zerofield}
\end{figure}

The power-broadened full-width-at-half-maximum of the ODMR peaks of this diamond was previously measured to be $\Gamma_0\approx0.6~{\rm MHz}$~\cite{SMI2025}, albeit under somewhat different microwave and laser powers. This value is consistent with our estimates of $\Gamma_0$. We also acquired ODMR spectra under zero applied field to infer $\Gamma_0$, Fig.~\ref{fig:zerofield}. The apparent FWHM is ${\sim}1.5~{\rm MHz}$, which is consistent with our estimates of $\Gamma_0$, but slightly higher, perhaps due to underlying strain-shift structure.

\section{\label{app:cal}Data collection and calibration}
\subsection{\label{app:save_file}Saved data}
Every experiment run through the LabVIEW interface saves a file containing the following information. For the ODMR spectrum recorded from the oscilloscope, we store the oscilloscope trace, trigger settings, number of averages, number of points in the trace, seconds per division, volts per division, and time delay. For the triaxial fluxgate magnetometer readings, we store the mean and standard deviation of each vector component, the number of samples, and the timestamp. For the three-channel current controller of the Helmholtz coils, we store the applied currents and voltages.  Additionally, each experiment saves metadata such as the timestamp of the experiment and the various calibration files that are most relevant for analysis.  

\subsection{\label{app:fit_timecalibration}Time-axis calibration}
Our magnetometer's accuracy relies on a precise mapping of microwave frequency to time on the oscilloscope. To correct for delays and nonlinearity in the microwave frequency ramp sweep, we use a mixer method for calibration. The signal from a microwave signal generator (TPI-1005-A) is set to a fixed frequency, either $2.62~{\rm GHz}$, $2.87~{\rm GHz}$ or $3.12~{\rm GHz}$, representing the beginning, middle, and end of an ODMR frequency ramp sweep, respectively. The fixed-frequency signal is fed into the radio-frequency (RF) port of a microwave frequency mixer (ZEM-4300MH+ mixer). The output of the microwave signal generator used for ODMR spectroscopy (SRS SG 384) is fed into the local oscillator (LO) port of the mixer. This microwave signal frequency is swept in the same way as in ODMR spectroscopy experiments, with a frequency span of $0.5~{\rm GHz}$. The intermediate frequency (IF) output of the mixer is sent to the oscilloscope (Yokagawa DL9040L). The oscilloscope trigger settings were kept the same as in ODMR spectroscopy experiments, including the time divisions of the scope, trigger position, and high-frequency trigger noise rejection. At the precise time when the RF port and LO port have the same frequency, the IF port output is at a zero-frequency null and is easy to localize in time. By alternating the fixed frequency sent to the RF port, we are able to locate the beginning, middle, and end of the ODMR frequency ramp sweep to within ${\lesssim}100~{\rm ns}$ (corresponding to $\lesssim100~{\rm Hz}$ in the microwave frequency). This time-frequency calibration was found to be linear and independent of the microwave frequency span. 

\subsection{\label{app:calibration_fghh}Fluxgate-Helmholtz calibration}
In order to apply magnetic fields of known amplitude and angle, calibration between the reference frames of the fluxgate and Helmholtz coils is needed. We chose the fluxgate axes to define the lab reference frame and attempted to align the Helmholtz coils along these axes. However, inevitably, there is some alignment error. Moreover, the lab ambient field needs to be determined in order to apply a known $\vec{B}_{\rm applied}$.

For this calibration, six measurements of the raw fluxgate field vector, $\vec{B}_{\rm fgraw}$, are made under the following applied currents. First, $+0.5~{\rm A}$ is applied to the $\hat{x'}$ coil pair, then $-0.5~{\rm A}$ is applied to the $\hat{x'}$ coil pair, and the process is repeated for the $\hat{y'}$ and $\hat{z'}$ coil pairs, respectively. A linear mapping of the form $\vec{B}_{\rm fgraw} = \mathcal{M} \vec{I} +\vec{b}$ is assumed, where $\vec{I}$ represents the current in the Helmholtz coils' Cartesian basis, $\vec{b}$ is the ambient field vector, and $\mathcal{M}$ is a $3\times3$ transformation matrix. The nine elements of $\mathcal{M}$ and three elements of $\vec{b}$ are extracted from the six sets of measurements of $\vec{B}_{\rm fgraw}$ and $\vec{I}$. This calibration procedure is repeated prior to obtaining the data analyzed in Figs.~\ref{fig:simul}(b), \ref{fig:results}, \ref{fig:deadzone}(a,b), and \ref{fig:stability} of the main text to account for any drift in the ambient field or variations in the Helmholtz coils alignment.
 
\subsection{\label{app:offset}Fluxgate-diamond calibration and offset field}
As discussed in Sec.~\ref{sec:angular} of the main text, determination of $\vec{B}_{\rm nv}$ uses independent measurements of the four NV axes ($\hat{e}_{\kappa}$) in the lab frame as well as a persistent offset field vector ($\vec{B}_{\rm off}$) between the fluxgate and diamond positions. The four NV symmetry axes are represented in matrix form in the diamond frame as:
\begin{equation}\label{eqn:nvs_standard}
{\rm NV_{std.}} = \pmat{1 & -1 & 1 & -1\\
1 & -1 & -1 & 1\\
1 & 1 & -1 & -1}.
\end{equation}
When the field is aligned with the NV axis, the secular Zeeman term in the ODMR transition frequencies, Eq.~\eqref{eqn:appxenergy} of the main text, is maximal, $\gamma_{\rm nv}|\vec{B}_{\rm nv}\cdot\hat{e}_{\kappa}|=\gamma_{\rm nv}|\vec{B}_{\rm nv}|$. Meanwhile the transitions frequencies of the other three NV axes overlap, with secular Zeeman term $\gamma_{\rm nv}|\vec{B}_{\rm nv}\cdot\hat{e}_{\kappa}|=\gamma_{\rm nv}|\vec{B}_{\rm nv}|/3$, owing to the $109.5\degree$ angle between $\vec{B}_{\rm nv}$ and $\hat{e}_{\kappa}$. This characteristic ODMR signature can be found through manual tuning or sweeping the field angle, enabling an initial estimate of the $\hat{e}_{\kappa}$ axes. Once two of the NV axes are found, the remaining two axes can, in principle, be inferred from symmetry. 

When we first performed this process, we found that the resulting estimates for the NV axes coordinates were not exactly $109.5\degree$ apart. Instead, the proper $C_{3v}$ symmetry was restored only if it was assumed that there was a static offset field, $\vec{B}_{\rm off}$, between the fluxgate and diamond positions. 

\begin{figure}
    \includegraphics[width=\columnwidth]{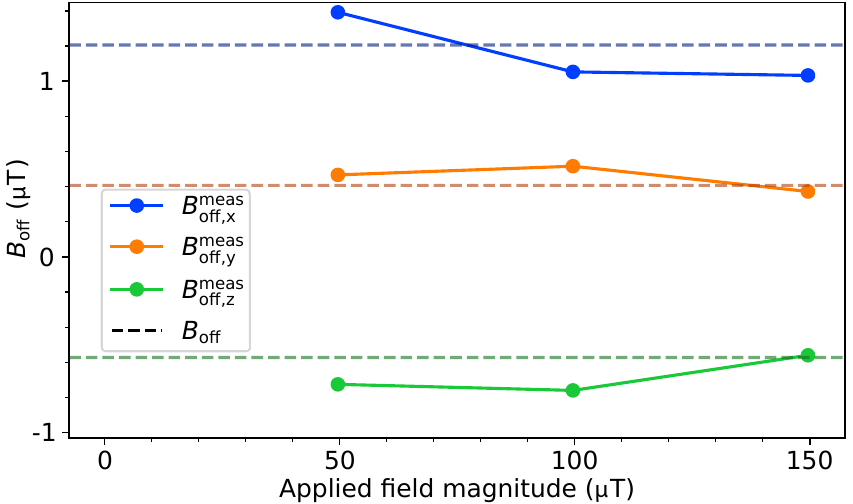}
    \caption{
    \textbf{Offset field vs. applied field magnitude.} For applied field magnitudes between 50 and $150~{\rm\upmu T}$, the field is applied along $\pm\hat{x}$, $\pm\hat{y}$, and $\pm\hat{z}$.  For each such pair, the offset field component is determined as described in the text.
    }
    \label{fig:offsetvB}
\end{figure}

To study this in more detail, we vary the magnitude of the applied field and determine the vector components of the offset field. For a given applied field magnitude, the magnetic field is applied along six field directions: $\pm\hat{x}$, $\pm\hat{y}$, and $\pm\hat{z}$. In each case, the applied field vector is measured using the raw, uncorrected fluxgate field, $\vec{B}_{\rm fgraw}=\vec{B}_{\rm fg}-\vec{B}_{\rm off}$. Thus, a non-zero $\vec{B}_{\rm off}$ means the applied field vector in the diamond position ($\vec{ B}_{\rm fg})$ is not exactly of constant magnitude for all six measurements. For each applied field, an ODMR spectrum is acquired and fit to determine the field magnitude within the diamond, $|\vec{B}_{\rm nv}|$. The offset field components are then estimated as:
\begin{equation}
B_{{\rm off},i}\approx \frac{B_{{\rm fgraw},i}^{+}\,B_{{\rm nv},i}^{-}\,-\,B_{{\rm fgraw},i}^{-}\,B_{{\rm nv},i}^{+}}{B_{{\rm nv},i}^{+}\,-\,B_{{\rm nv},i}^{-}},
\end{equation} where $B_{{\rm nv},i}^{\pm}$ is the field magnitude within the diamond for an applied field along $\pm\hat{i}$ and $B_{{\rm fgraw},i}^{\pm}$ are the corresponding fluxgate raw values. As shown in Fig.~\ref{fig:offsetvB}, the estimated $B_{{\rm off},i}$ components are largely independent of applied field amplitude. This indicates that the source of $\vec{B}_{\rm off}$ is a static gradient, as opposed to the result of some magnetizable object in the apparatus. 

\begin{figure}[hbt]
    \includegraphics[width=\columnwidth]{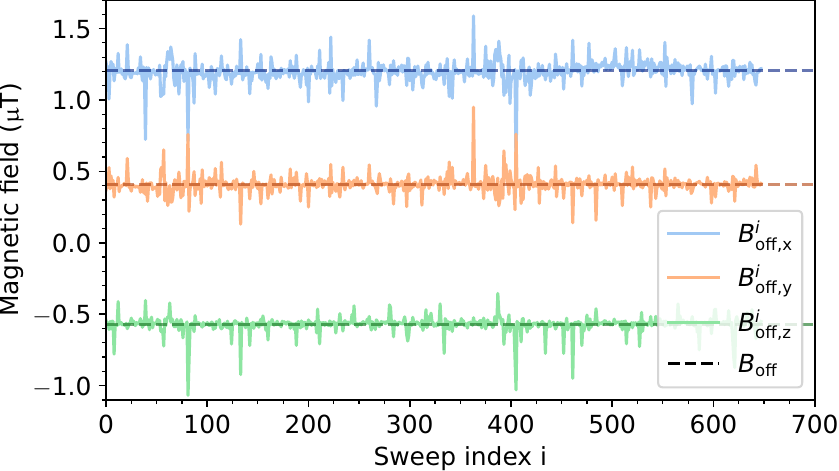}
    \caption{
    \textbf{Offset field vs. applied field angles.} The applied field of $50~{\rm \upmu T}$ is applied for 1296 field angles, corresponding to positive and negative complements along 648 axes. The offset field components are computed for each axis (denoted as ``Sweep index, i'') and are found to be relatively constant. 
    }
    \label{fig:offsetsweep}
\end{figure}

The constant nature of $\vec{B}_{\rm off}$ is further confirmed by experiments where we hold the applied field magnitude constant and vary the field angle, as was done for Fig.~\ref{fig:results} of the main text. For these data, we conduct a more complete analysis, using the matrix formalism for $\mathcal{E}$, to determine the components of $\vec{B}_{\rm off}$. Moreover, now there are $648$ different axes where antiparallel fields are applied. Figure~\ref{fig:offsetsweep} shows the inferred components of $\vec{B}_{\rm off}$ at each applied field axis in the sweep. The estimated offset field is largely independent of angle. The results of this analysis are that the offset field is consistent with $\vec{B}_{\rm off} = [1.20, 0.42, -0.57]~{\rm\upmu T}$. This is also consistent with the residual $\vec{B}_{\rm fgraw}$ observed when tuning the NV ODMR spectrum to $\vec{B}_{\rm nv}=0$, Fig.~\ref{fig:zerofield}.

With the fluxgate offset calibrated, we then found the precise positions of the NV axes in the lab frame by fitting for their coordinates. Specifically, using the data set in Fig.~\ref{fig:results} of the main text, we selected the NV axes coordinates that minimized the mean of the angle offset errors, $\bar{\Delta\alpha}$, while preserving $C_{3v}$ symmetry. The NV axes in the fluxgate frame were thus determined to be:
\begin{equation}\label{eqn:nvs_actual}
{\rm NV_{expt.}}{=}\pmat{0.478 & 0.537 & -0.896 & -0.119\\
0.693 & -0.843 & -0.181 & 0.331\\
-0.541 & 0.011 & -0.406 & 0.936}.
\end{equation}

\section{\label{app:fit}Fitting the vector magnetic field}
\subsection{\label{app:fit_hamiltonian}Hamiltonian-constrained fits}
The simplest fit to an ODMR spectrum of an ensemble of $^{14}$NV centers would have independent parameters to describe the FWHM, amplitude, and frequency of each of the 24 peaks, resulting in at least 72 fit parameters. However, the central frequencies would then need to be used to find the components of $\vec{B}_{\rm nv}$, and in some cases this may lead to a poor match due to fitting ambiguities. Instead, we constrain the central frequencies to be an allowed set of transition frequencies given by diagonalization of the Hamiltonian in Eq.~\eqref{eqn:hamiltonian} of the main text.

In our fitting routine, the amplitude and FWHM of each of the 24 Lorentzian peaks are independent fit parameters, but the central frequencies are constrained by 9 parameters of the Hamiltonian: $B_{{\rm nv},x}$, $B_{{\rm nv},y}$, $B_{{\rm nv},z}$, $D=2870.08\,\pm\,0.04~{\rm MHz}$, $E=1~{\rm MHz}$, $A_{\parallel}=-2.17~{\rm MHz}$, $P=-4.95~{\rm MHz}$, $\gamma_{\rm nv}=28.03~{\rm GHz/T}$, and $\gamma_{\rm nuc}=-3.07~{\rm MHz/T}$. All parameters are fixed as constants, based on known values, except $E$ which is independently measured (see, for example, Fig.~\ref{fig:zerofield}), $D$ which is allowed to vary by $\pm40~{\rm kHz}$ to account for temperature drifts, and $\vec{B}_{\rm nv}$. We neglect transverse nuclear Zeeman and transverse hyperfine coupling terms, as they were found to have negligible contribution ($\ll1~{\rm kHz}$) to the transition frequencies for the low fields studied here.  

In our implementation, the Hamiltonian of a single NV axis is written as:
\begin{align}
H &= H_S\otimes\ro 1+H_{SI}+\ro 1\otimes H_I,\label{eqn:nv_hamiltonian}\\
H_S &= \ro S\cdot \mathcal{D}\cdot\ro S + \ro S\cdot\mathcal{G_{\rm s}}\cdot\ro B\notag\\
H_{SI} &= \left(\ro S\otimes\ro 1\right)\cdot\mathcal{A}\cdot\left(\ro 1\otimes\ro I\right)\notag\\
H_I &= \ro I\cdot\mathcal{P}\cdot\ro I + \ro I\cdot\mathcal{G_{\rm i}}\cdot\ro B\notag 
\end{align}
where $\ro I = \pmat{I_x&I_y&I_z}$ and $\ro S=\pmat{S_x&S_y&S_z}$ are Pauli matrices for the spin-1 nuclear and electron spin vectors, respectively and $\ro B$ is the magnetic field vector inside the diamond illumination volume. The remaining variables are treated as diagonal matrices with diagonal entries given by:
\begin{align*}
    \mathcal{D} &= \pmat{E&-E&D},
    & \mathcal{G_{\rm s}} &= \pmat{\gamma_{\rm nv}& \gamma_{\rm nv}& \gamma_{\rm nv}}, &
    \\
    \mathcal{A} &= \pmat{0&&0&&A_{\parallel}}, &
    \\
     \mathcal{P} &= \pmat{0&&0&&P},
     & \mathcal{G_{\rm i}} &= \pmat{ 0&& 0&& \gamma_{\rm nuc}}. &
\end{align*} 
As discussed in \ref{app:offset}, in order to express the magnetic field $\vec{B}$ as $\vec{B}_{\rm nv}$ in the lab frame, a rotation matrix is independently determined and applied to the Zeeman terms in Expressions~\eqref{eqn:nv_hamiltonian}. Obtaining a complete set of 24 spin transition frequencies involves applying the appropriate rotation to each of the four NV axes, diagonalizing the Hamiltonian of each NV axis, and evaluating the results (as described later in this section) to determine the six transition frequencies of each NV axis. 

The eigenvalues and eigenvectors of the Hamiltonian are obtained with numpy's eigh function, which is a relatively fast numerical routine that assumes a Hermitian Hamiltonian. In calculations, the eigenvectors $\ket{\xi^k}$ are expressed as a vector of the coefficients, $a_{m_i,m_s}$, of states $\ket{m_S,m_I}$:
\[\ket{\xi^k} = \pmat{a_{1,1}\\a_{1,0}\\a_{1,-1}\\a_{0,1}\\a_{0,0}\\a_{0,-1}\\a_{-1,1}\\a_{-1,0}\\a_{-1,-1}}.\]
The spin transition selection rules require: $\Delta m_s=\pm1,~\Delta m_i=0$.
We express the selection rules as a 9x9 matrix: 
\[M=\pmat{0&1&0\\1&0&1\\0&1&0}\otimes\pmat{1&0&0\\0&1&0\\0&0&1}\]
The selection rules are then incorporated by evaluating the transition probabilities:
\begin{equation}\label{eq:transitions}
\mathcal M_{ij}=\left|\bracket{\xi^i}M{\xi^j}\right|^2.
\end{equation}
For a given eigenvector $\ket{\xi^j}$, the associated second eigenvector that maximizes $\mathcal{M}_{ij}$ is defined as the ``allowed'' transition. The transition frequency is defined as the difference in their associated eigenvalues.

For efficiency, the process of identifying allowed transition frequencies is vectorized.  First, a 9x9 matrix of all possible transition frequencies $\mathcal{V}_{ij}=\xi^i-\xi^j$ is created (whether they are allowed transitions or not). The three eigenvectors associated with the $m_s=0$ states are identified and labeled by their associated $m_i$ value as $\ket{\xi^{-1}}$, $\ket{\xi^{0}}$, and $\ket{\xi^{1}}$.  For the remaining six eigenvectors, their allowed transition is identified by applying Eq.~\eqref{eq:transitions}, with $\bra{\xi^j}=\bra{\xi^{\{-1,0,1\}}}$, and selecting the $m_s=0$ eigenvector that maximizes the transition probability. The corresponding transition frequencies in $\mathcal{V}$ are identified as the six allowed transition frequencies. 

In the fit routine, the 24 Lorentzian peak amplitudes are normalized by a variable representing the average amplitude of all peaks. This normalization allows us to separate the relative variation in peak amplitudes (due to, for example, the distinct contrast of each NV axis related to ODMR polarization anisotropy) from the overall amplitude (due to, for example, changes in laser power) to ease manual fitting.  The FWHM of each peak is treated in a similar manner.  With the baseline $h$ representing the fluorescence photodetector voltage off resonance, this leaves a total of 53 fit variables: 24 amplitudes, 24 FWHM linewidths, 3 components of $\vec{B}_{\rm nv}$, $D$, and $h$.

\subsection{\label{app:fit_statistics}Fit initial parameters and bounds}
To get ODMR spectrum fits to converge quickly, we use knowledge of prior ODMR spectra as well as the fluxgate $\vec{B}_{\rm fg}$ to apply initial parameters and bounds. Specifically, when fitting the ODMR spectra presented in Fig.~\ref{fig:results} and Fig.~\ref{fig:stability} of the main text, the initial parameters of the ODMR peak amplitudes, linewidths, and $\vec{B}_{\rm nv}$ fit variables for the first spectrum are drawn from previous fits under similar conditions. After the first fit, initial parameters and bounds are adjusted based on the fit results of the ODMR spectrum obtained at the most similar field angle.  

With this algorithm, the vast majority of fits converge, and the fitted $\vec{B}_{\rm nv}$ components are continuous with their neighbors (in angle space, for Fig.~\ref{fig:results}, or time, for Fig.~\ref{fig:stability}). For a few percent of fits in Fig.~\ref{fig:results}, there is substantial spectral overlap and the fit either does not converge to the requested tolerance or the resulting fit values are discontinuous with their neighbors. In these cases, we use initial parameters that are the average of the two nearest neighbors (in angle space) with successful fits.

As discussed in Sec.~\ref{sec:experiment} of the main text (see also \ref{app:temperature}), $D$ is allowed to vary within $2870.08\pm0.04~{\rm kHz}$ due to its temperature dependence.  The $E$ parameter, if left as a free fit parameter, can vary substantially with no noticeable effect on $\vec{B}_{\rm nv}$ unless the applied field is approximately orthogonal to one of the NV axes.  Thus, we use $E=1~{\rm MHz}$ as a fixed parameter. This estimate for $E$ is based on fit results from ODMR spectra obtained at applied field angles that are orthogonal to one of the NV axes and at zero field, see Fig.~\ref{fig:zerofield}. 

\section{\label{app:solidangle}Solid angle weighing}
When extracting statistics from the angular dependence and accuracy data in Sec.~\ref{sec:angular}, and the estimation theory analysis in Sec.~\ref{sec:crb}, of the main text, we corrected for nonuniform sampling. The underlying data were acquired by incrementing $\theta$ and $\phi$ in constant increments. While convenient, this leads to non-uniform coverage over the solid angle in field vector space. For example, at $\theta=0,\pi$, the field angle is the same regardless of $\phi$, resulting in numerous redundant measurements.

Our correction involves weighting each measurement by the solid angle it occupies relative to neighboring measurements. For an angular sweep with angular steps $d\theta$ in the polar angle and $d\phi$ in the azimuthal angle, the weighting function is given by:
\begin{equation}\label{eqn:solidangle}
W(\theta, \phi; d\theta, d\phi) = d\phi\left[ 
\cos\left(\theta-d\theta/2\right)
- \cos\left(\theta+d\theta/2\right)
\right],
\end{equation}
where $\theta\pm d\theta/2$ is clipped to [0, $\pi$] in the cosine argument. Note that this weighting function is approximately proportional to $\sin{\theta}$ except for the edge cases when $\theta=0,\pi$. 

The weighted mean, $\mu$, and standard deviation, $\sigma$, of a parameter $X$ sampled at polar and azimuthal angles $\left\{\theta_i\right\}$ and $\left\{\phi_j\right\}$ in constant increments is calculated as:
\begin{align}
    \mu(X) &= \frac{\sum_{ij} W\pP{\theta_i, \phi_j} X\pP{\theta_i, \phi_j}}{\sum_{ij} W\pP{\theta_i, \phi_j}},\\
    \sigma(X) &= \frac{\sum_{ij} W\pP{\theta_i, \phi_j} \left[ X\pP{\theta_i, \phi_j} - \mu(X) \right]^2}{\sum_{ij} W\pP{\theta_i, \phi_j}}.
\end{align}
These formulas are applied when computing statistics of the various metrics characterizing the angular dependence, $\sigma({\epsilon})$, $\mu({\Delta\alpha})$, and $\sigma({\Delta\vec{B}})$, as well as determining the fraction of the total solid angle free from deadzones in the estimation theory analysis.

\section{\label{app:fit_enhtensor}Modeling Enhancement}
In Sec.~\ref{sec:angular} of the main text, we introduce a $3\times3$ matrix form of the enhancement, $\mathcal{E}$, to account for the observed slight anisotropicity of the flux concentrator. Specifically, the field vector in the diamond illumination region, $\vec{B}_{\rm nv}$, is related to the fluxgate-recorded applied field vector, $\vec{B}_{\rm fg}$, by:
\begin{equation}
\label{eq:enhmatrix}
\vec{B}_{\rm nv} = \mathcal{E} (\vec{B}_{\rm fgraw}+\vec{B}_{\rm off})=\mathcal{E} \vec{B}_{\rm fg},
\end{equation}
where $\vec{B}_{\rm fgraw}$ is the raw fluxgate magnetometer reading and $\vec{B}_{\rm off}$ is the persistent offset field vector between fluxgate and diamond positions (\ref{app:offset}). This simple model for the enhancement is supported by finite-element magnetostatic simulations, where Eq.~\eqref{eq:enhmatrix} accurately describes the behavior of ferrite cones with angle and displacement errors.

When the enhancement matrix is pre-calibrated, the resulting field estimation is $\mathcal{E}^{-1}\vec{B}_{\rm fg}$, which results in a more accurate field estimation compared to the assumption of a scalar (perfectly isotropic) enhancement factor. We use three different methods to estimate $\mathcal{E}$. The simplest method assumes that $\vec{B}_{\rm off}$ has been independently measured and requires acquisition of just three ODMR spectra at different applied field vectors. In this case, application of Eq.~\eqref{eq:enhmatrix} results in nine equations (one for each field component of each measurement), with nine unknowns (the different elements of $\mathcal{E}$), allowing for estimation of the full $\mathcal{E}$ matrix. 

By adding a fourth ODMR spectrum acquired at a different applied field, the components of $\vec{B}_{\rm off}$ can be uniquely estimated, as application of Eq.~\eqref{eq:enhmatrix} results results in 12 equations and 12 unknowns. In practice, we instead apply six measurements to allow for a slightly more precise estimate. In this case, the components of $\mathcal{E}$ and $\vec{B}_{\rm off}$ can be estimated by minimizing the residuals function: $\sum_{i=1}^6 |\mathcal{E}^{-1}\vec{B}_{\rm nv}-(\vec{B}_{\rm fgraw}+\vec{B}_{\rm off})|^2$, where $i$ iterates through the six measurements. This was the method used to determine statistics for the matrix-corrected values of $\bar{\Delta\alpha}$, $\sigma(\Delta\vec{B})$, and $\sigma(|\Delta\vec{B}|/|\vec{B}_{\rm fg}|)$ in Sec.~\ref{sec:angular} of the main text.

\begin{figure}[bt]
    \includegraphics[width=\columnwidth]{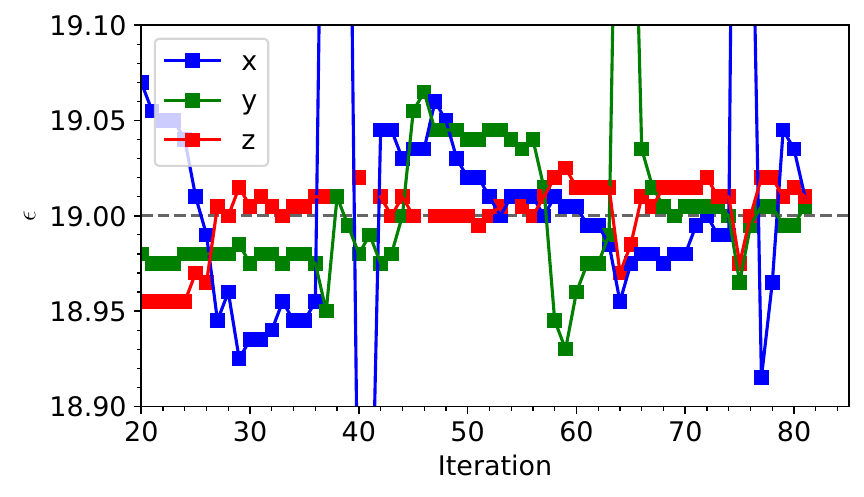}
    \caption{\textbf{Tuning to isotropicity.} The adjustable cones are manually translated using the fine-thread screws and the diagonal elements of $\mathcal{E}$ are plotted based on a six-measurement calibration. For most iterations, only a single cone was adjusted. When the diagonal elements are equal and close to $19$, the anisotropicity of the flux concentrator is minimized and further analysis is conducted.}
    \label{fig:enhtuning}
\end{figure}

\begin{figure*}[hbt]
    \includegraphics[width=.9\textwidth]{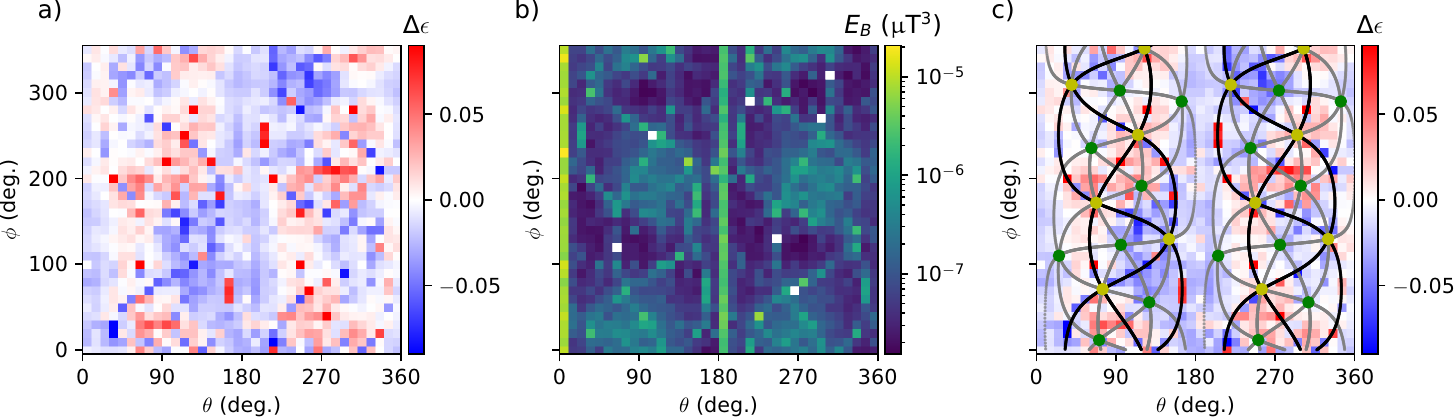}
    \caption{
    \textbf{Enhancement linearity.} 
    (a) Enhancement errors, $\Delta\epsilon=\epsilon_{\rm expt}-\epsilon_{\rm model}$, as a function of applied field angle, $(\theta,\phi)$. The weighted standard deviation is $\sigma({\Delta\epsilon})=0.024$, corresponding to a relative standard deviation $\sigma({\Delta\epsilon})/\bar{\epsilon}=1.2\times10^{-3}$. (b) The determinant of the covariance matrix from ODMR fits, $E_B$, as a function of $(\theta,\phi)$. Discontinuities are present for similar field angles as in (a). (c) Illustration of the field angles corresponding to \{111\} (yellow) and \{100\} (green) directions, along with regions with singly degenerate ODMR peaks (gray) and doubly-degenerate peaks (black), overlayed on (a).  The doubly-degenerate ODMR peak regions correlate with regions of larger model errors and increased fit uncertainty.
    }
    \label{fig:linearenh}
\end{figure*}

These rapid methods of pre-calibration are used to position the adjustable cones. We turn the adjustable fine-thread screws of each of the three cones. For each position, we apply fields approximately along $\pm\hat{x}$, $\pm\hat{y}$, and $\pm\hat{z}$ and record and fit ODMR spectra. From these data, we compute $\mathcal{E}$. Figure~\ref{fig:enhtuning} shows the diagonal elements of $\mathcal{E}$ for different adjustments of the cones. We observed experimentally (and confirmed through magnetostatic modeling) that when the diagonal elements are equal and close to $19$, the anisotropicity of the flux concentrator is minimized. Another property of $\mathcal{E}$ that is useful for tuning the adjustable cones is the difference in extreme values of $\epsilon(\theta,\phi)$. To compute this difference, we use the singular value decomposition of $\mathcal{E}$ obtained by the six-measurement method.

The final method of calibration uses a set of 1296 ODMR spectra obtained by incrementing $\theta$ and $\phi$ by $10\degree$, as done for Fig.~\ref{fig:results} of the main text. Similar to the six-measurement calibration, the estimation of $\mathcal{E}$ is done by minimizing residuals. For the spectra in Fig.~\ref{fig:results} of the main text, the best-fit enhancement matrix is:
\begin{equation}
    \mathcal{E} = \pmat{
    19.135 & 0.159 & -0.013\\
    0.127 & 18.985 & 0.187\\
    -0.208 & 0.240 & 19.038}.\label{eqn:numerenh}
\end{equation}

Figure~\ref{fig:linearenh}(a) shows the angular dependence of the enhancement errors, $\Delta\epsilon=\epsilon_{\rm expt}-\epsilon_{\rm model}$, when comparing the experimentally-observed enhancement factor, $\epsilon_{\rm expt}=|\vec{B}_{\rm nv}|/|\vec{B}_{\rm fg}|$, with that estimated by the enhancement matrix, $\epsilon_{\rm model}=|\mathcal{E}\vec{B}_{\rm fg}|/|\vec{B}_{\rm fg}|$, where Eq.~\eqref{eqn:numerenh} is used for $\mathcal{E}$. The values of $\Delta \epsilon$ are generally small for all field angles, indicating good agreement, though some fixed patterns are present indicative of model imperfections. Notably, the largest deviations in Fig.~\ref{fig:linearenh}(a) form a spiraling pattern at special field angles. A similar, but even more stark, pattern is observed when plotting the determinant of the ODMR fit covariance matrix, $E_B$, as seen in Fig.~\ref{fig:linearenh}(b). This indicates that much of the deviation is due to systematic errors in fitting the ODMR spectra. It was found that these errors occur when multiple ODMR peaks overlap. Figure~\ref{fig:linearenh}(c) shows the angles corresponding to multiple ODMR peak degeneracies overlayed onto the data in Fig.~\ref{fig:linearenh}(a). It is seen that the largest deviations in model estimate occur when ODMR peaks overlap.

\section{\label{app:fisher}Cram\'er-Rao bound analysis}
In Sec.~\ref{sec:crb} of the main text, we apply a Cram\'er-Rao lower bound analysis~\cite{IRE2005} to estimate the minimum attainable uncertainty in the magnetic field components $B_x, B_y, B_z$ from ODMR spectroscopy and quantify deadzones. When a microwave frequency $f$ is applied, the ODMR signal, expressed as the number of photons detected over the detection interval, is $N_f$. The signal's probability density function is denoted $p_{f_i}(N_{f_i}\vert \vec{B})$, where $\vec{B}$ is the ambient field measured by the vector magnetometer (in the main text, this variable is $\vec{B}_{\rm nv}$). We assume that $N_f$ is measured for discrete, uniformly spaced values of $f_i$ ranging from $i=1$ to $i_{\rm max}$. We can then define the log-likelihood function as:
\begin{equation}
\label{eq:loglikely}
    \ln\left[L(\vec{N}\vert\vec{B})\right] = \sum_{i=1}^{i_{\rm max}}\ln \left[p_{f_i}(N_{f_i}\vert\vec{B})\right].
\end{equation}

\begin{figure*}[hbt]
    \includegraphics[width=.75\textwidth]{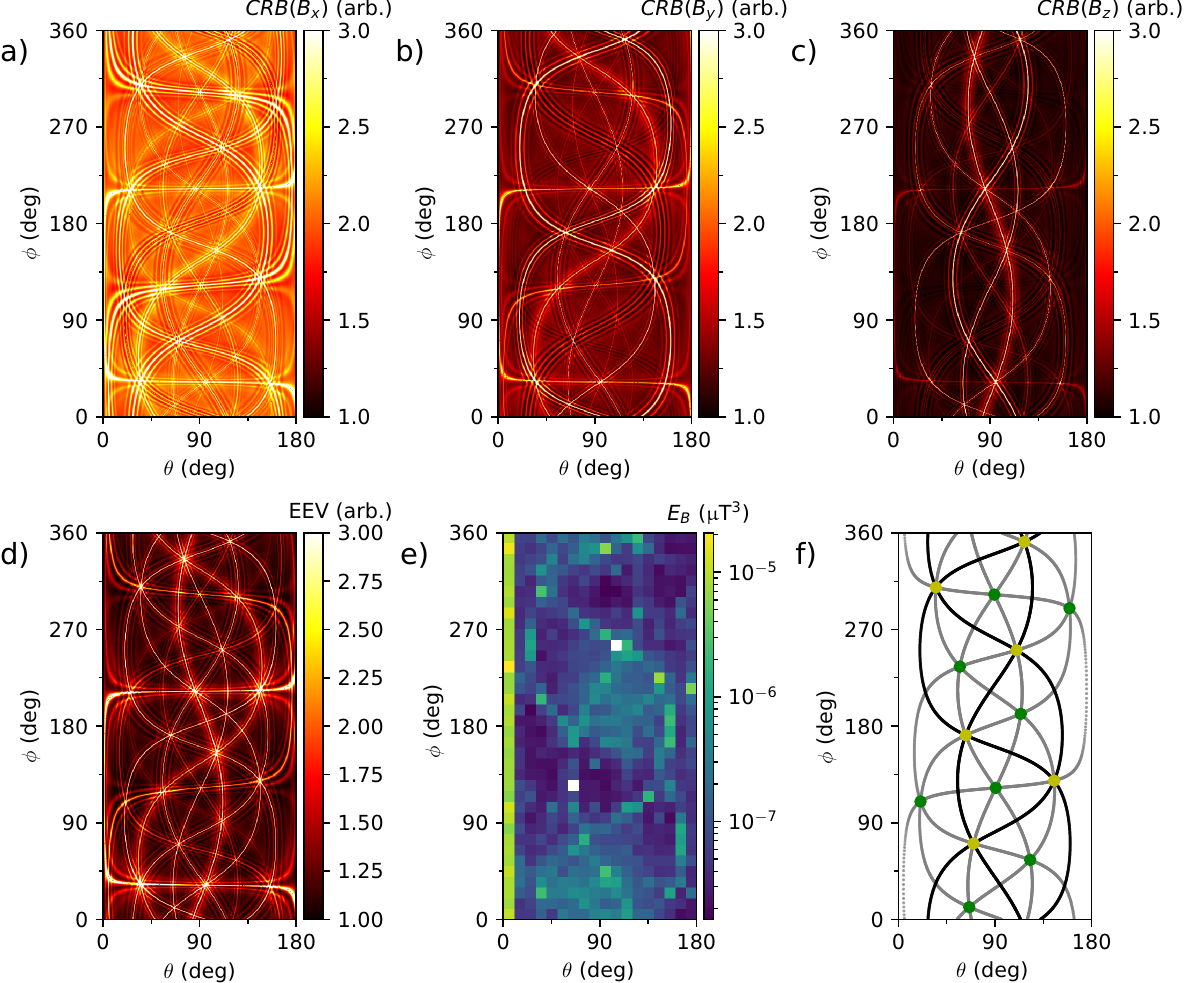}
    \caption{\textbf{Full deadzone analysis.} (a) Cram{\`e}r-Rao lower bound for estimation uncertainty of $B_x$, calculated for $|\vec{B}_{\rm nv}|=1~{\rm mT}$ over the entire $4\pi$ solid angle. (b,c) Same as (a) except for $B_y$ and $B_z$, respectively. (d) Error ellipsoid volume, EEV, as a function of field angle. (e) The determinant of the covariance matrix from ODMR fits of the data in in Fig.~\ref{fig:results} as a function of $(\theta,\phi)$. 
    (f) Illustration of the field angles corresponding to \{111\} (yellow) and \{100\} (green) directions, along with regions with singly degenerate ODMR peaks (gray) and doubly-degenerate peaks (black).}
    \label{fig:deadzone_appendix}
\end{figure*}

We next define the Fisher information matrix $\mathcal{I}$ as:
\begin{equation}
    \mathcal{I}_{p,q}=-E\left(\frac{\partial^2\ln\left[ L(\vec{N}\vert {\vec{B})}\right]}{\partial B_{p}\partial B_{q}}\right),
    \label{eq:Fisher_ij}
\end{equation}
where $E$ denotes the expectation value with respect to the applied field. Assuming that the photon count probability density function $p_f$ is Poissonian \cite{IRE2005}, the above expression simplifies to:
\begin{equation}
  \mathcal{I}_{p,q}=\sum_{i=1}^{i_{\rm max}}\frac{1}{F_{f_i}(\vec{B})}\frac{\partial F_{f_i}(\vec{B})}{\partial B_p}\frac{\partial F_{f_i}(\vec{B})}{\partial B_q},
   \label{eq:fisher} 
\end{equation}
where $F_{f_i}(\vec{B})$ is the expectation value of the signal $N_{f_i}$. For the NV vector magnetometer, the term $1/F_{f_{i}}(\vec{B})$ is approximately a constant, since the fluorescence level only varies by a few percent over the spectrum, see Fig.~\ref{fig:simul}(b). We note that if the probability density function $p_f$ is instead governed by white Gaussian noise of standard deviation $s$, the term $1/F_{f_i}(\vec{B})$ in Eq.~\eqref{eq:fisher} is replaced by $1/s^2\,{\approx}\,{\rm const}$. In either case, Eq.~\eqref{eq:fisher} simplifies to:
\begin{equation}
\label{eq:fisherapp}     
\mathcal{I}_{p,q}\propto\sum_{i=1}^{i_{\rm max}}\frac{\partial F_{f_i}(\vec{B})}{\partial B_p}\frac{\partial F_{f_i}(\vec{B})}{\partial B_q}.
\end{equation}

The Cram\'er Rao inequality states that the lower bound of the variance of field component $B_p$ is equal to the corresponding diagonal matrix element of the inverse Fisher information matrix:
\begin{equation}
\label{eq:crineq}
    {\rm min}[\sigma^2(B_p)] = [\mathcal{I}^{-1}]_{pp}.
\end{equation}
The Cram\'er Rao bound for estimating parameter $B_p$ is then defined as the square root of the minimum variance:
\begin{equation}
\label{eq:crbapp}
\mathrm{CRB}(B_p)=\sqrt{[\mathcal{I}^{-1}]_{pp}},
\end{equation}
as discussed in the main text.

We use the Cram\'er Rao bound analysis to compare how the uncertainty in field-component estimation changes with respect to the field angle. For this analysis, we do not assume the presence of flux concentrators, we merely take the applied field to be $1~{\rm mT}$, to reflect the effect of flux concentration on Earth's field. We evaluate the partial derivatives in Expression~\eqref{eq:crbapp} numerically. For a given applied field vector, $\vec{B}$, the expected ODMR response function, $F_{f_i}(\vec{B})$ does not have a simple analytical form, but it is known from the NV Hamiltonian and the experimentally-observed ODMR lineshapes. The 24 ODMR frequencies are computed as described in \ref{app:fit_hamiltonian}, and the lineshapes are taken to be Lorentzians, with amplitudes and linewidths for each NV axis given as the average experimental values. We then evaluate the partial derivative of $F_{f_i}(\vec{B_p})$ with respect to field component $B_x$ numerically as:
\begin{equation}
    \frac{\partial F_{f_i} (\vec{B})}{\partial B_x} \approx \frac{F_{f_i}(B_x+ \delta B_x, B_y, B_z) - F_{f_i}(B_x, B_y, B_z)}{\delta B_x}. 
\end{equation}
The other partial derivatives are computed in a similar manner, with $\delta B_x = \delta B_y = \delta B_z = 50~{\rm nT}$. The Cram\'er-Rao lower bounds are then computed numerically using Eq.~\eqref{eq:crbapp}. Our choices of $\delta B_{x,y,z}$ produce spectral shifts that are ${\sim}4$ orders of magnitude smaller than the linewidth and are thus assumed to be sufficiently small. However, automatic differentiation algorithms may be preferable for more challenging calculations.

Figure~\ref{fig:deadzone_appendix}(a-c) shows the Cram\'er-Rao lower bounds of each field component as a function of field angle. All values of $\mathrm{CRB}(B_p)$ are normalized to the average value observed at field angles when all 24 ODMR peaks are well resolved. There is a slight asymmetry in that $\mathrm{CRB}(B_x)$ values are generally slightly larger than $\mathrm{CRB}(B_y)$ values, which in turn are slightly larger than $\mathrm{CRB}(B_z)$ values. We tentatively attribute this to our choice of microwave and optical polarizations, which may result in larger ODMR amplitudes for NV axes with greater $\hat{z}$ projection (and smaller amplitudes for NV axes with greater $\hat{x}$ projection). More prominently, there are certain applied field angles where the Cram\'er-Rao lower bound features large peaks, indicating a loss of precision in the magnetic field estimates. These regions coincide with field angles where the EEV is large, Fig.~\ref{fig:deadzone_appendix}(d), the fit uncertainty is large, Fig.~\ref{fig:deadzone_appendix}(e), and multiple ODMR peaks overlap, Fig.~\ref{fig:deadzone_appendix}(f). The interpretation is that, for these field angles of ODMR overlap, a small change in magnetic field causes very little change in the ODMR spectrum.  

For some range of applied field angles, the $\mathrm{CRB}(B_p)$ is so large that the magnetometer precision has degraded to an unacceptable degree, which we label "deadzones". We define the threshold for a deadzone to be angles where $\mathrm{CRB}(B_p)$ is a factor of 2 larger than that of the fully resolved case. The fractions of the full solid angle considered to be deadzones for each of the field components are summarized in table~\ref{tbl:scrape_deadzones}. The areas are calculated by counting the number of points (pairs of $(\theta,\phi)$) where $\mathrm{CRB}(B_p)$ is above the cut-off threshold and weighting them as described in \ref{app:solidangle}.

\begin{table}[hbt]
    \begin{tabular}{c | c}
      Field component~ & ~Deadzone fraction (\%)\\
      \hline
      $\mathrm{CRB}(B_x)$ & 1.74\\
      $\mathrm{CRB}(B_y)$ & 2.12\\
      $\mathrm{CRB}(B_z)$ & 2.29\\
    \end{tabular}
    \caption{\textbf{Deadzone fractions.} Calculated values of $\mathrm{CRB}(B_p)$ as a function of field angle, as presented in Fig.~\ref{fig:deadzone_appendix}(a-c), are processed to determine the deadzone fraction. The deadzone fraction is defined as the fraction of the full $4\pi$ solid angle where $\mathrm{CRB}(B_p)$ is a factor of 2 larger than that of the fully resolved case.}
    \label{tbl:fullsweepdead}
\end{table}

\section{\label{app:deadzonescrape}Deadzones in the literature}
In Sec.~\ref{sec:crb} of the main text, we compare the deadzone fraction of our device to two other works: Ref.~\cite{BEN2010} which explored the angular dependence of different modes of an alkali-metal vapor magnetometer and Ref.~\cite{GE2020} which analyzed different modes of an Overhauser nuclear magnetic resonance magnetometer. These works were selected because: i) they reported on bias-field-free spin resonance magnetometers, and ii) they presented enough data on the angular response that we could estimate the deadzone fraction using our Cram\'er-Rao Bound metric (\ref{app:fisher}). Nevertheless, we did not use the raw data, instead extrapolating them from figures, and we resorted to making a number of simplifying assumptions in order to extract a deadzone fraction. Thus, our estimates should be considered as very rough. Moreover, these works reported on scalar, total-field magnetometers; care must be taken to compare with the full-vector magnetometer studied here.

For the alkali-metal vapor optical magnetometer~\cite{BEN2010}, the magnetic field is inferred from the position of a type of ODMR peak, similar to this work. We thus make an assumption that the noise in the magnetometer signal is dominated by photon shot noise. However, unlike in our case, the off-resonance optical signal is much smaller than the on-resonance signal, so an assumption of the noise being independent of amplitude would be inappropriate. Instead, we assume the noise variance scales linearly with ODMR peak amplitude such that $\mathrm{CRB}(B)$ increases by a factor of two when the amplitude falls by a factor of four. Table~\ref{tbl:scrape_deadzones} shows the resulting rough estimates of the deadzone fraction for the two different modalities studied in that work. We estimate a deadzone fraction ranging from $1\%$ to $35\%$, depending on the mode of operation.

For the Overhauser nuclear magnetic resonance magnetometer~\cite{GE2020}, we assume that the noise is dominated by white Johnson noise. In this case, the noise is independent of amplitude, and $\mathrm{CRB}(B)$ is assumed to increase linearly with signal amplitude. Table~\ref{tbl:scrape_deadzones} shows the resulting rough estimates of the deadzone fraction for the two different modalities studied in that work as well as a theoretical expression. We estimate a deadzone fraction ranging from $10\%$ to $50\%$, depending on the mode of operation.

\begin{table}[hbt]
    \begin{tabular}{c | c | c}
      Source       & Criterion     & Deadzone fraction (\%)\\
      \hline
      \cite{BEN2010} - top plot    & $<$ max/4 & 1\\
      \cite{BEN2010} - bottom plot & $<$ max/4 & 35\\
      \cite{GE2020} - Eq 2              & $<$ max/2 & 50\\
      \cite{GE2020} - Eq 3              & $<$ max/2 & 40\\
      \cite{GE2020} - Experimental data & $<$ max/2 & 10
    \end{tabular}
    \caption{\textbf{Deadzone estimates for two prior works.} Rough calculated deadzone fractions extracted for an alkali-metal vapor magnetometer work~\cite{BEN2010} and Overhauser nuclear magnetic resonance magnetometer~\cite{GE2020}. The deadzone fraction is defined as the fraction of the full $4\pi$ solid angle where the expected $\mathrm{CRB}(B)$ is a factor of 2 larger than the ideal case. A number of assumptions are made (see text), so the values should be treated as rough estimates.}
    \label{tbl:scrape_deadzones}
\end{table}

\section{\label{app:temperature}Ambient temperature variation}

\begin{figure}[hbt]
    \includegraphics[width=\columnwidth]{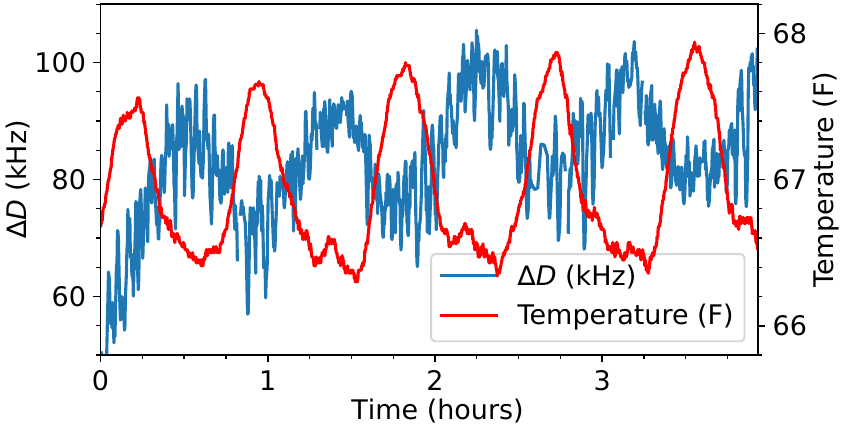}
    \caption{\textbf{Ambient temperature variation.} The fitted zero-field splitting parameter, given as $\Delta D=D-2870~{\rm MHz}$, is plotted as a function of time. The lab temperature, $T$, in Fahrenheit is also plotted as a function of time. Note that these two data sets were not obtained simultaneously, but rather on separate nights. Nevertheless, both $\Delta D$ and $T$ oscillate with a ${\sim}50$-minute period.}
    \label{fig:thermalD}
\end{figure} 

As discussed in Sec.~\ref{sec:temporal}, and shown in Fig.~\ref{fig:stability}(e), the fitted value of the zero-field splitting parameter, $D$, was found to oscillate in a periodic fashion over time. The variation is attributed to fluctuation in the lab ambient temperature. To verify this, we recorded ODMR spectra overnight and fit them to observe the time dependence of $D$. On a separate night, we recorded the lab temperature as a function of time. Both time traces are shown in Fig.~\ref{fig:thermalD}. The traces are found to oscillate with a period of ${\sim}50~{\rm minutes}$. The regular oscillation is likely related to the temperature feedback control of the room.

\section{\label{app:simulation}Magnetostatic modeling of flux-concentrator anisotropy}
The slight anisotropy of our flux concentrator, as recorded in Fig.~\ref{fig:results} of the main text, indicates some deviation from the ideal geometry. In order to gain an intuition for what sorts of deviations could replicate the behavior, we used magnetostatic modeling of different device geometries.  

We conducted an optimization to find a device geometry that most closely reproduces the angular dependence of the enhancement factor, $\epsilon(\theta,\phi)$, observed in Fig.~\ref{fig:results}(b). We allowed the cone azimuthal and polar angles to vary, as well as their individual positions, and the sensor position. The optimization was performed through the \texttt{scikit-optimize} python module's \texttt{forest\_minimize} algorithm, which uses a tree based regression model.  The model builds the geometry, evaluates the simulation, and extracts parameters and average field over the spherical sensor volume through the \texttt{MPh} python wrapper. To address the large parameter space and nontrivial optimization landscape, the best guess was iteratively improved, and bounds were made tighter and tighter as the algorithm converged. 

\begin{figure}[hbt] 
\includegraphics[width=\columnwidth]{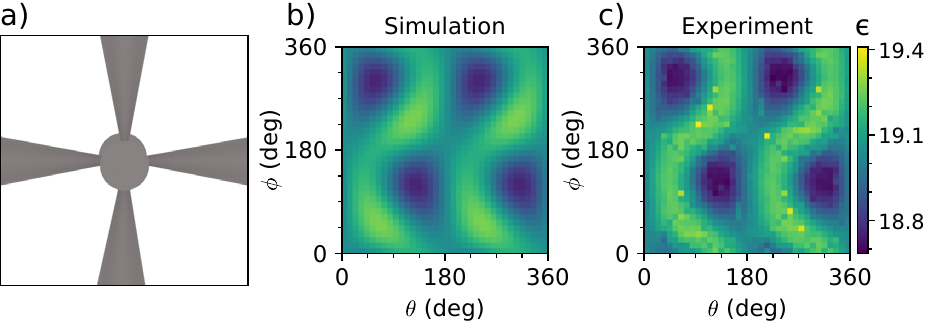}
    \caption{\textbf{Simulating experimental anisotropy.} (a) Geometry resulting from optimization to match simulated $\epsilon(\theta,\phi)$ to that of experiment.  Horizontal and vertical axes are $\hat{x}$ and $\hat{z}$ directions, respectively. (b) Simulated map of $\epsilon(\theta,\phi)$.  (c) Experimental map of $\epsilon(\theta,\phi)$, identical to Fig.~\ref{fig:results}(b) of the main text.  
    \label{fig:comsol_replicate}
    }
\end{figure}

\begin{table}[hbt]
    \begin{tabular}{c | c}
      Parameter & Value(s)\\
      \hline

        Cone tip-to-tip starting gap &                0.634 mm\\
         Sensor displacement &  [19.37, 40.5, 0.87] $\rm\upmu m$\\
        X-cone pair rotation & -0.54$\degree$$\hat{z}$, then 2.20$\degree$$\hat{y}$ \\
        Y-cone pair rotation & -0.69$\degree$$\hat{z}$, then -0.56$\degree$$\hat{x}$ \\
        Z-cone pair rotation & 2.05$\degree$$\hat{x}$, then 1.82$\degree$$\hat{y}$ \\
            +X cone rotation & -1.760$\degree$$\hat{z}$, then -0.431$\degree$$\hat{y}$ \\
            +Y cone rotation & -1.373$\degree$$\hat{z}$, then -0.942$\degree$$\hat{x}$ \\
            +Z cone rotation & 0.191$\degree$$\hat{x}$, then 2.188$\degree$$\hat{y}$ \\
            -X cone rotation & -1.037$\degree$$\hat{z}$, then 1.383$\degree$$\hat{y}$ \\
            -Y cone rotation & 0.332$\degree$$\hat{z}$, then 1.739$\degree$$\hat{x}$ \\
            -Z cone rotation & -0.652$\degree$$\hat{x}$, then 1.053$\degree$$\hat{y}$ \\
        +x cone displacement & [20.61, 18.38, -6.43] $\rm\upmu m$\\
        -x cone displacement &  [46.94, 5.48, 8.67] $\rm\upmu m$\\
        +y cone displacement & [22.02, -18.98, 35.88] $\rm\upmu m$\\
        -y cone displacement & [-1.33, -21.22, 9.4] $\rm\upmu m$\\
        +z cone displacement & [-1.87, 8.35, 42.17] $\rm\upmu m$\\
        -z cone displacement & [18.44, -9.98, 53.38] $\rm\upmu m$\\
    \end{tabular}
    \caption{\textbf{Simulation parameters.} Best-fit geometry parameters used in the simulation in Fig.~\ref{fig:comsol_replicate}. 
    }
    \label{tbl:simulexptparams}
\end{table}

The resulting geometry is illustrated in Fig.~\ref{fig:comsol_replicate}(a), with best-fit parameters listed in Table~\ref{tbl:simulexptparams}. The simulated map of $\epsilon(\theta,\phi)$ is shown in Fig.~\ref{fig:comsol_replicate}(b) and is seen to be in good agreement with that observed experimentally, Fig.~\ref{fig:comsol_replicate}(c). The match between simulation and experiment indicates that the experimental device can be simulated by finite-element magnetostatic modeling. However, we expect that a number of different geometries, including degrees of freedom not explored in this simulation, could replicate the experimental behavior. The geometry found in our optimization routine is interpreted as just one possible example.

\newpage

%

\end{document}